\documentclass[preprint]{revtex4}

\usepackage{custom_defn}
\usepackage{graphicx} 
\usepackage{dcolumn} 
\usepackage{amssymb}
\usepackage{amsmath}
\usepackage{bm}
\usepackage[dvips]{color}
\usepackage{times}
\usepackage{graphicx}
\usepackage{ulem}
\usepackage{subfigure}
\usepackage{tabularx}
\usepackage{feynmp} \unitlength=1mm
\usepackage{slashed}
\usepackage{slashbox}

\usepackage{amstext}
\usepackage{array}

\usepackage{bbm}

\usepackage[colorlinks=true
,urlcolor=blue
,anchorcolor=blue
,citecolor=red
,filecolor=blue
,linkcolor=blue
,menucolor=blue
,pagecolor=blue
,linktocpage=true
,pdfproducer=medialab
]{hyperref}

\allowdisplaybreaks 

\newcommand{\bqa}{\begin{eqnarray}} 
\newcommand{\eqa}{\end{eqnarray}}
\newcommand{\nn}{\nonumber \\}

\definecolor{new_color}{RGB}{50,155,0}

\newcommand{\ep}{\epsilon}
\newcommand{\beq}{\begin{equation}}
\newcommand{\eeq}{\end{equation}}
\newcommand{\beqq}{\begin{equation*}}
\newcommand{\eeqq}{\end{equation*}}
\newcommand\beqa{\begin{eqnarray}}
\newcommand\eeqa{\end{eqnarray}}

\newcommand{\f}{\frac}

\def\s{\sigma}
\newcommand{\eps}{\varepsilon}

\def\g{\gamma}

\begin{document}

\title{
Recent Developments in Non-Fermi Liquid Theory
}

\author{Sung-Sik Lee\\
\vspace{0.3cm}
{\normalsize{$^1$Department of Physics $\&$ Astronomy, 
McMaster University,}}\\
{\normalsize{1280 Main St. W., Hamilton ON L8S 4M1, Canada}}
\vspace{0.2cm}\\
{\normalsize{$^2$Perimeter Institute for Theoretical 
Physics,}}\\
{\normalsize{31 Caroline St. N., Waterloo ON N2L 2Y5, 
Canada}}
}

\date{\today}

\begin{abstract}

Non-Fermi liquids arise when metals 
are subject to singular interactions 
mediated by soft collective modes.
In the absence of well-defined quasiparticle,
universal physics of non-Fermi liquids 
is captured by interacting field theories
which replace Landau Fermi liquid theory.
In this review, we discuss two approaches
that have been recently developed
for non-Fermi liquid theory
with emphasis on two space dimensions.
The first is a perturbative scheme 
based on a dimensional regularization,
which achieves a controlled access to the low-energy physics
by tuning the number of co-dimensions of Fermi surface.
The second is a non-perturbative approach 
which treats the interaction ahead of the kinetic term
through a non-Gaussian scaling called interaction-driven scaling.
Examples of strongly coupled non-Fermi liquids
amenable to exact treatments 
through the interaction-driven scaling 
are discussed.

\end{abstract}

\maketitle


\section{Introduction}

Metal is among the most delicate
states of quantum matter.
Due to the presence of extensive gapless modes
that support long-range entanglement in real space,
metals are highly susceptible to external perturbations.
In view of this,  
it is rather remarkable   
that Fermi liquid metals
are more or less immune to quantum fluctuations
generated by the screened Coulomb interaction.
In Fermi liquids, 
many-body eigenstates of interacting fermions
are still labeled by the occupation numbers of 
quasiparticles\cite{LFL}.
The existence of 
well defined single-particle excitations
is attributed to the Pauli exclusion principle 
that severely limits non-forward scatterings near the Fermi surface\cite{1992hep.th...10046P,RevModPhys.66.129}.

With the discovery of 
high-temperature superconductors,
heavy fermion compounds
and iron pnictides,
it became clear that 
quantum criticalities provide natural routes
to non-Fermi liquids
which lie beyond the quasiparticle paradigm.
At quantum critical points,
non-Fermi liquids  
are realized
as soft order parameter fluctuations
mediate singular interactions between electrons\cite{PhysRevB.14.1165,PhysRevB.48.7183,RevModPhys.79.1015,RevModPhys.73.797}. 
Although one generally needs a fine tuning 
to reach a critical point in the zero temperature limit,
the underlying non-Fermi liquid state
can dictate the universal scaling behaviors\cite{PhysRevB.78.035103} 
over an extended region of the phase diagram at finite temperatures. 
If a collective mode is  protected dynamically not by a fine-tuning,
non-Fermi liquid states can be realized as a phase without a fine tuning.

Arguably, non-Fermi liquids in layered systems are most interesting
from theoretical perspective.
Typically being below the upper critical dimension,
non-Fermi liquids in two dimensions are expected 
to deviate strongly from the Fermi liquids. 
On the other hand, metals in two dimensions 
can support an extended Fermi surface 
unlike in one dimension.
The combination of strong infrared quantum fluctuations
and the presence of extended manifolds of gapless modes
makes non-Fermi liquids in two dimensions rather unique\cite{
PhysRevB.8.2649,
PhysRevB.40.11571,
PhysRevB.50.14048,
PhysRevB.50.17917,
PhysRevLett.63.680,
polchinski1994low,
PhysRevB.46.5621,
nayak1994non,
PhysRevB.80.165102,
PhysRevB.82.075127,
PhysRevB.82.045121,
jiang2013non,
PhysRevB.88.245106,
PhysRevB.90.045121,
PhysRevB.92.041112}.
Besides dimensionality,
another important factor that
determines universal properties of non-Fermi liquids
is the wavevector, $\vec Q$ 
carried by the soft collective mode at zero energy.
For $\vec Q=0$, low-energy collective modes induce small-angle scatterings
everywhere on the Fermi surface.
Examples include 
the nematic critical point
and the U(1) spin liquid with spinon Fermi surface.
In this case, the low-energy theory 
describes a `hot Fermi surface'.
Non-Fermi liquids with nonzero $\vec Q$ can arise
near quantum critical points associated with
density wave transitions
of spin, charge or orbital.
With $\vec Q \neq 0$,
fermions can interact with low-energy collective modes
only near a sub-manifold of the Fermi surface
connected by $\vec Q$.
In two dimensions, the sub-manifold is a discrete set of `hot spots'.

The first part of the review
covers non-Fermi liquids with hot Fermi surfaces.
In Sec. II A, we introduce the theory
that describes the non-Fermi liquid realized 
at the Ising-nematic quantum critical point 
in two dimensions.
Among the schemes that have been proposed 
for the purpose of gaining a controlled access 
to the physics of non-Fermi liquid,
we focus on a dimensional regularization scheme
which tunes the number of co-dimensions of Fermi surface.
Although the dimensional regularization scheme allows
one to understand non-Fermi liquids reliably near the upper critical dimension, 
the perturbative approach has the fundamental limit 
when applied to the strongly coupled theory in two dimensions.
In Sec. II B, a non-perturbative approach 
which treats interactions ahead of the kinetic term is discussed.
In particular, we introduce the notion of interaction-driven scaling,
and discuss an example of non-Fermi liquids 
whose exact critical exponents 
can be obtained from the interaction-driven scaling. 
The second half of the review covers 
the antiferromagnetic quantum critical metal
as an example of hot-spot theories.
In Sec. III A, the perturbative results obtained from 
the dimensional regularization scheme are discussed. 
In Sec. III B, we discuss the non-perturbative solution 
for the theory in two dimensions, which gives exact critical exponents
through an Ansatz constructed from the interaction-driven scaling.

\section{Theory of hot Fermi surface}

The Ising-nematic phase transition
refers to a spontaneous breaking of
the four-fold rotational symmetry 
to the two-fold symmetry.
A positive or negative order parameter
represents one of the two symmetry breaking patterns.
If a metal undergoes a nematic quantum phase transition,
a non-Fermi liquid arises at the critical point\cite{PhysRevB.64.195109,PhysRevLett.91.066402,PhysRevB.75.033304,PhysRevB.74.195126,PhysRevB.81.045110,PhysRevB.80.165116,PhysRevB.82.075127,PhysRevB.88.245106}.
The Ising order parameter is strongly damped by particle-hole excitations
while fermions near the Fermi surface undergo persistent scatterings 
by the fluctuating order parameter.
The theoretical goal is to capture the universal properties 
that result from the interplay 
between the gapless collective mode
and soft fluctuations of the Fermi surface.

\begin{figure}
\begin{center}
\subfigure[]{
\includegraphics[scale=0.30]{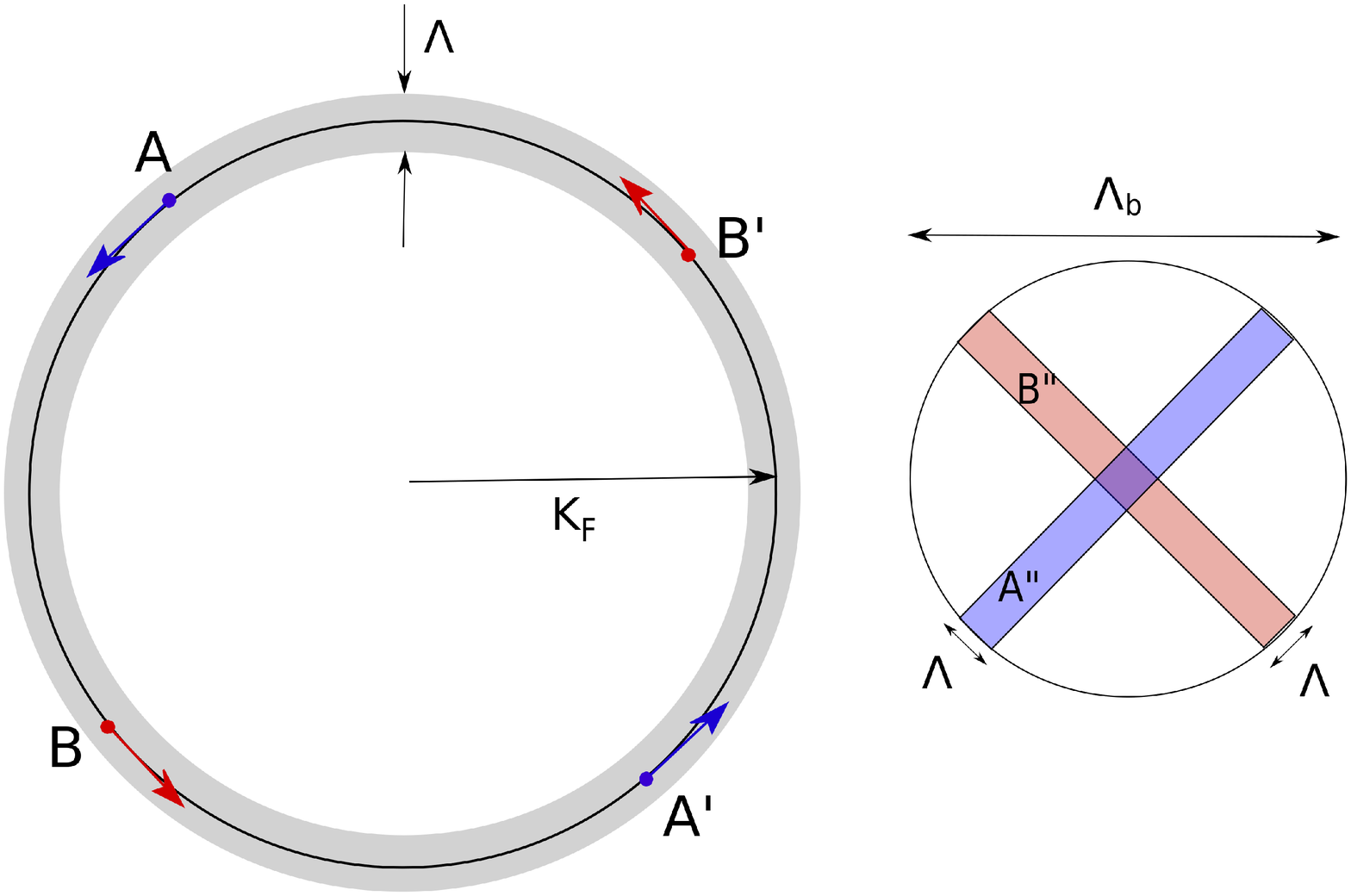}
\label{fig:locality}
}
~~~~~~~~
\subfigure[]{
\includegraphics[scale=0.35]{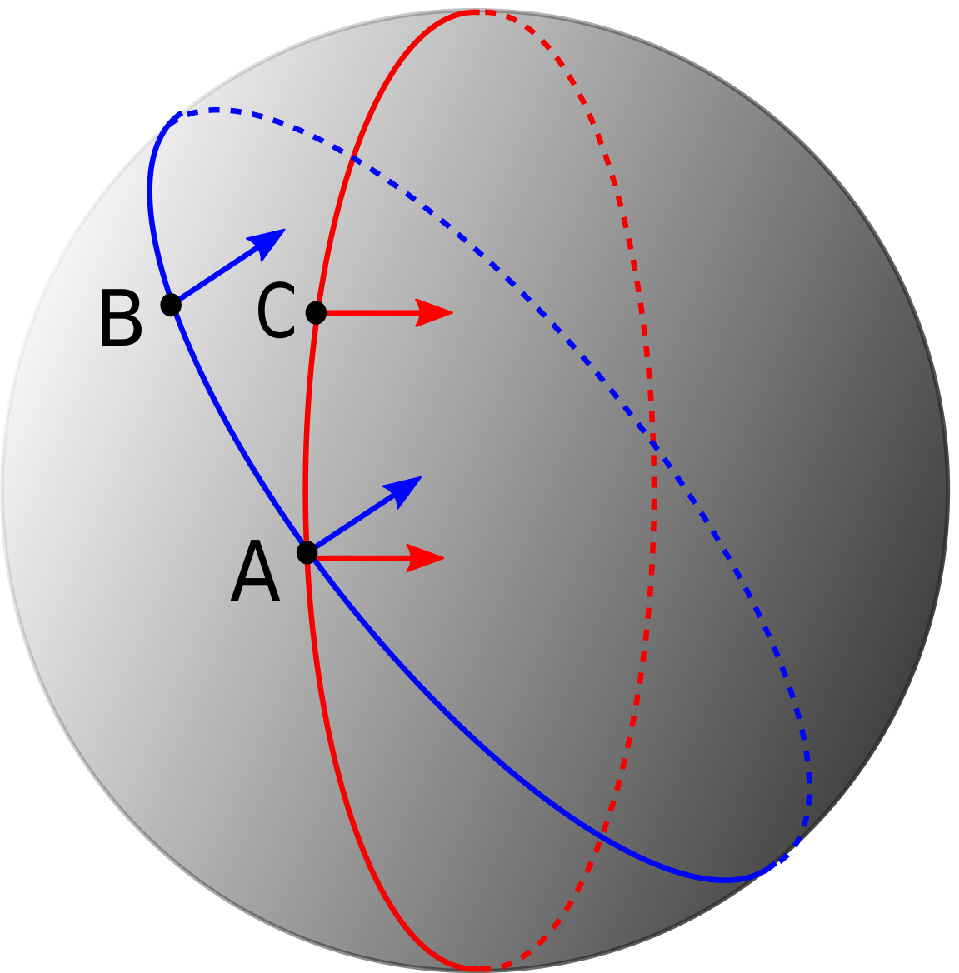}
\label{fig:2dFS}
}
\end{center}
\caption{
(a) The origin of the emergent locality in momentum space.
The left figure represents a Fermi surface with size $K_F$.
The shaded region is the phase space for low-energy fermionic excitations 
with an energy cut-off $\Lambda \ll K_F$.
The right figure denotes the momentum space of a collective mode.
The momentum cut-off for the collective mode
$\Lambda_b \sim \sqrt{ K_F \Lambda}$
is much larger than $\Lambda$
because a fermion can absorb a boson with momentum much larger than $\Lambda$
if the momentum of the boson is parallel to the Fermi surface.
When Fermi surface is one-dimensional,
there exists only a discrete set of points 
whose tangent vectors are parallel or anti-parallel to each other.
For example, point $A$ ($B$) share the anti-parallel tangent vector 
only with point $A'$ ($B'$). 
Therefore the fermions near $A$ and $A'$ ($B$ and $B'$)
couple only with the boson in region $A''$ ($B''$) at low energies.
Since the overlap between $A''$ and $B''$ becomes vanishingly small in the low energy limit,
the sector that describes $\{A, A', A''\}$ decouples
from that of $\{B, B', B''\}$.
(b) 
The locality in momentum space is lost 
as soon as the dimension of Fermi surface becomes greater than one.
When the dimension of Fermi surface is greater than one with co-dimension one,
one can find a common tangent vector for any two points on the Fermi surface.
For example,  the point $A$ have tangent vectors
that are tangent to point $B$ and $C$ 
on a two-dimensional Fermi surface.
}
\end{figure}


In two dimensions,
the nematic order parameter is coupled to the 
quadrupolar distortion of the Fermi surface,
$H_{int} = \sum_{\vec K,  \vec q}~ (\cos K_x  - \cos K_y ) \phi(\vec q) c_{j}^\dagger(\vec K+\vec q) c_j(\vec K)$,
where $\phi(\vec q)$ is the collective mode for the nematic order parameter 
with momentum $\vec q$,
and $c_{j}(\vec K)$ is the electron field with momentum $\vec K$ and spin $j=\uparrow, \downarrow$.   
The fermion at momentum ${\vec K}$ on the Fermi surface 
is mainly coupled with the collective mode
whose momentum is tangential to the Fermi surface at ${\vec K}$.
This is because fermions can absorb or emit bosons
whose momenta are tangential to the Fermi surface
while staying close to the Fermi surface.
Because fermions from different patches of Fermi surface
interact with the boson with largely disjoint sets of momenta,
the inter-patch coupling is small in the low energy limit, 
unless the Fermi surfaces in two patches 
are locally parallel or anti-parallel\cite{polchinski1994low,PhysRevB.78.085129}.
This is illustrated in \fig{fig:locality}.
The exception that breaks the locality in momentum space is 
the short-range four-fermion interaction 
in the pairing channel.
We will discuss the issue of superconductivity 
at the end of the section.

The  locality in momentum space allows one to 
decompose the full theory into a sum of two-patch theories.
Each two-patch theory describes electronic excitations near two antipodal points
and the collective mode whose momentum is close to be tangential to the Fermi surface
at the antipodal points.
Let us the consider the two-patch theory
which describes the patches
centered at ${\vec K} = \pm K_F \hat x$. 
The action is written as 
\bqa
S & = &  \sum_{s=\pm} \sum_{j=\uparrow, \downarrow} \int \frac{d^3k}{(2\pi)^{3}}  
\psi_{s,j}^\dagger (k)
\Bigl[ 
 i k_0   +  s  k_x +  k_y^2  \Bigr] \psi_{s,j}(k) \nonumber \\
 &+& \frac{1}{2} \int  \frac{d^3q}{(2\pi)^{3}} 
 \left[ q_0^2 + c_x^2  q_x^2 +  c_y^2 q_y^2 ) \right] | \phi(q)|^2 \nonumber \\
 &+&  g_0 \sum_{s=\pm} \sum_{j=\uparrow, \downarrow}  
\int \frac{d^3k d^3q}{(2\pi)^{6}}   ~ \phi(q) ~  \psi^\dagger_{s,j}(k+q) \psi_{s,j}(k).
\label{act0}
\eqa
Here $k=(k_0,k_x,k_y)$ denotes the frequency and momentum in the Euclidean spacetime.
$\psi_{+,j}(k) = c_j(k_0, K_F \hat x + \vec k)$
and $\psi_{-,j}(k)= c_j(k_0, -K_F \hat x + \vec k)$
represent the right and left moving fermion with spin $j$ respectively.
The momentum for the fermion field has been shifted 
such that $\vec k=0$ represents the point
on the Fermi surface in each patch.
$k_x, k_y$ have been rescaled independently 
so that the absolute value of Fermi velocity
and the local curvature of the Fermi surface become one. 
$g_0$ is the fermion-boson coupling,
and $(c_x, c_y)$ is the velocity of the collective mode.
Theories for other types of non-Fermi liquids with hot Fermi surface
are similar to \eq{act0}.
The only major modification is the nature of the vertex in the coupling.
One exception is when the Fermi surface is not locally parabolic\cite{PhysRevB.90.045121}.
For example, non-parabolic patches arise near inflection points on Fermi surface,
which can be viewed as a `multi-critical' theory 
tuned by the angle around the Fermi surface.

One of the most important aspects of metal
is the fact that there are infinitely many gapless modes
near the Fermi surface.
One way to formalize this is to view
the Fermi surface in two dimensions 
as an infinite set of one-dimensional Dirac fermions,
where the momentum along the Fermi surface labels 
the continuously many gapless modes.
To make this precise, 
the right and left moving fermions can be combined into a spinor, 
$\Psi_j(k)^T = \left( 
\psi_{+,j}(k),
\psi_{-,j}^\dagger(-k)
\right)$.
In this representation, 
the fermion kinetic term becomes
$
S_F  =   \sum_{j} \int \frac{d^3k}{(2\pi)^{3}}  
\bar \Psi_j(k)
\Bigl[ 
 i k_0 \gamma_0  + i (   k_x +  k_y^2 )  \gamma_1 \Bigr] 
\Psi_{j}(k)$,
where $\gamma_0 = \sigma_y$, $\gamma_1 = \sigma_x$ 
are the gamma matrices for the two component spinor,
and $\bar \Psi \equiv \Psi^\dagger \gamma_0$.
Because the Fermi surface is locally parabolic,
the scaling dimension of $k_y$ is 
half the dimension of $k_{x}$.
Under the tree-level scaling that leaves
the kinetic terms invariant, 
only $  q_{y}^2  |\phi(q)|^2$ 
is marginal for the boson kinetic term,
and  $( q_0^2 + c_x^2 q_{x}^2 ) | \phi(q)|^2$ can be dropped.
Physically, this implies that the dynamics of the boson
is strongly dressed by particle-hole excitations 
to the extent that some parts of the bare kinetic term 
become unimportant at low energies.

\subsection{Perturbative approach}
\label{HotSurfacenonPerturb}

At the non-interacting fixed point,
the coupling has the scaling dimension $1/2$, 
and grows as energy is lowered.
Given that there is no general non-perturbative method
for strongly coupled theories,
it is natural to start by deforming the theory so that
the effect of quantum fluctuations can be included perturbatively.

Over the years, different theoretical schemes have been proposed to this end.
The most straightforward deformation 
is to enlarge the number of spin components to a large number $j=1,2,...,N$\cite
{polchinski1994low,PhysRevB.50.14048,PhysRevB.50.17917}.
In relativistic quantum field theories, 
the mean-field theory is applicable in the large $N$ limit,
and quantum fluctuations can be perturbatively
included order by order in the $1/N$ expansion.
However, this is no longer the case in the presence of Fermi surface.
Due to the abundance of soft particle-hole excitation near the Fermi surface,
low-energy quantum fluctuations are not completely tamed even in the large $N$ limit.
The angle around the Fermi surface effectively becomes an additional flavor, 
which turns the theory to be like a matrix theory. 
The resulting theory is not solvable even in the large $N$ limit 
due to the proliferation of planar graphs\cite{PhysRevB.80.165102}.
In the one-patch theory that describes non-Fermi liquids without time-reversal and parity invariance,
the $1/N$ expansion is organized by the genus of Feynman diagrams.
For non-chiral theories, even non-planar diagrams play an important role\cite{PhysRevB.82.075127}.
Understanding the nature of 
non-Fermi liquids in the large $N$ limit 
remains as an open problem. 
As alternatives, 
the limits in which the number of fermionic species 
is much smaller than the number of bosonic species 
have been also considered\cite{PhysRevB.50.14048,2016arXiv161205326S,PhysRevB.89.165114}.

Another deformation scheme is a dynamical tuning,
where one tunes the bare dispersion of excitations\cite{nayak1994non,PhysRevB.82.045121}.
For example, one can change the kinetic term of 
the collective mode to $|\vec q|^{1+\ep} |\phi(q)|^2$,
where $\ep$ is a tuning parameter.
As $\ep$ decreases from $1$, 
the density of state for the collective mode 
is reduced at low energies,
and one can use $\ep$ as a control parameter to tame quantum fluctuations.
This scheme has the merit of keeping a finite density of state of fermion and preserving all microscopic symmetries.
A downside is that it breaks the locality in real space,
which blocks the collective mode 
from acquiring an anomalous dimension.

Finally, dimensional regularization schemes are considered.
Unlike relativistic field theories,
both the space dimension $d$ and the dimension of Fermi surface $d_f$ 
can be tuned independently\cite{PhysRevB.92.035141}.
Among the infinitely many ways to approach 
the original theory with $(d, d_f)=(2,1)$,
the most intuitive extension is probably to keep the co-dimension ($d-d_f$)
of the Fermi surface fixed\cite{PhysRevLett.74.1423,PhysRevB.88.125116}.
Under such an extension,
the fermion kinetic term is generalized to
$S_F  =   \sum_{j} \int \frac{d^{d+1}k}{(2\pi)^{d+1}}  
\bar \Psi_j(k)
\Bigl[ 
i k_0 \gamma_0  + i (   k_1 +  \sum_{\mu=2}^d k_\mu^2  )  \gamma_1 \Bigr] 
\Psi_{j}(k)$
in $d$ space dimensions.
While this has the advantage of keeping a non-zero density of states at zero energy,
it has a drawback of
spoiling the emergent locality 
in momentum space (not in real space).
For $d_f>1$,
any two points on the Fermi surface share a common tangential vector
as is illustrated in \fig{fig:2dFS},
and fermions remain strongly coupled across the entire Fermi surface.
Since the size of Fermi surface enters
as a relevant scale in all low energy observables,
a UV/IR mixing arises\cite{PhysRevB.92.035141}.

The dimensional regularization scheme 
that avoids the UV/IR mixing is the one in which
the manifold of the gapless modes is unchanged
as the number of dimensions is tuned
\cite{PhysRevLett.102.046406,PhysRevB.88.245106,PhysRevB.91.125136}.
%
%
Although it seems unnatural, 
the deformation which tunes the number of co-dimensions
is in line with the original idea of dimensional regularization scheme
applied to relativistic quantum field theories,
where the number of gapless points in momentum space
is not increased while the space dimension is increased.
One takes advantage of the reduced densities of state 
to control quantum fluctuations at low energies.
More concretely, one extends the theory in two dimensions 
to a theory that describes a line node in general dimensions 
whose kinetic energy reads
$S_F  =   i \sum_{j} \int \frac{d^{d+1} k}{(2\pi)^{d+1}}  
\bar \Psi_j(k)
\Bigl[ 
\sum_{\mu=0}^{d-2}  k_\mu \gamma_\mu + (   k_{d-1} +  k_d^2 )  \gamma_2 \Bigr] 
\Psi_{j}(k)$,
where $k_1, .., k_{d-2}$ are the newly added directions
which are transverse to the Fermi surface,
and $(k_{d-1},k_d)$ represents the original two-dimensional plane
in which the line node is located.
In three dimensions, the theory describes
a $p$-wave superconductor with a line node.
This scheme has the advantage of keeping
the locality in real space and 
the emergent locality in momentum space.
The drawback of this scheme is 
to break certain symmetry of the original model
in gapping out the parts of the Fermi surface away from the line nodes.
In this case, the global $U(1)$ 
associated with the transformation
$\Psi \rightarrow e^{i \theta \sigma_3} \Psi$
is broken down to $Z_2$ by pairing.
While the full symmetry can be kept if one gives up locality in real space\cite{PhysRevLett.102.046406},
one has to pay the price of breaking some symmetry
if one keeps the locality in real space\cite{PhysRevB.88.245106,PhysRevB.91.125136}.

Among the deformation schemes discussed above, 
we employ the latter dimensional regularization scheme 
which puts more emphasis on localities (both in real and momentum spaces)
than symmetry.
The theory that continuously interpolates
the Fermi surface in two dimensions
to the line node in three dimensions reads\cite{PhysRevB.88.245106}
\begin{eqnarray}
\label{act4}
S & = &  \sum_{j} \int \frac{d^{d+1}k}{(2\pi)^{d+1}} \bar \Psi_j(k)
\Bigl[ 
i {\bf \Gamma} \cdot {\bf  K} + i \gamma_{d-1} \delta_k \Bigr] \Psi_{j}(k) 
~ + 
\frac{1}{2} \int  \frac{d^{d+1}q}{(2\pi)^{d+1}}
~ q_y^2  \phi(-q) \phi(q) 
\nonumber\\
&+&  \frac{i \sqrt{d-1}}{\sqrt{N}}  g \mu^{\frac{5-2d}{4}} \sum_{j}  
\int \frac{d^{d+1}k d^{d+1}q}{(2\pi)^{2d+2}}  ~
\phi(q) \bar \Psi_{j}(k+q) \gamma_{d-1} \Psi_{j}(k).
\end{eqnarray}
Here the spin index is generalized to $j=1,2,..,N$.
${\bf  K} \equiv (k_0, k_1,\ldots, k_{d-2})$
represents frequency and $(d-2)$ components of  
the full $(d+1)$-dimensional energy-momentum vector.
$\delta_k =  k_{x}+ \sqrt{d-1} k_y^2$
is the energy dispersion of the fermion
in the two-dimensional subspace, 
where we keep the notation $k_x, k_y$ in favor of $k_{d-1}, k_d$.
The gamma matrices associated with ${\bf K}$ 
are written as
${\bf  \Gamma} \equiv (\gamma_0, \gamma_1,\ldots, \gamma_{d-2})$.
Since the space dimension of 
interest lies between $2$ and $3$,
the number of spinor components is fixed to be two. 
The Fermi surface is located 
at $k_1 = .. = k_{d-2} = 0$ and $\delta_k=0$.
The $(d-1)$ constraints for $d$ components of momentum
gives a one-dimensional manifold 
embedded in the $d$-dimensional momentum space.
In \eq{act4}, $c_y$ has been absorbed into the field redefinition.
$\mu$ is an energy scale introduced to make $g$ dimensionless.
The upper critical dimension of the theory is $d_c = 5/2$,
and the non-Fermi liquid state can be accessed perturbatively
in $d=5/2 - \ep$ for small $\ep$.

In computing the quantum effective action that determines physical observables,
there is one peculiar feature
that is not present in relativistic field theories.
Because the minimal action does not include 
the full kinetic term for the collective mode,
the boson propagator needs to be dressed with self-energy
before loops that include internal boson lines are computed.
At the one-loop order, the boson propagator becomes
\bqa
D_1(k) = 
\frac{1}{ k_y^2  + \beta_d g^2 \mu^{\epsilon} \frac{| {\bf K}|^{d-1}}{|k_y|} },
\label{babos}
\eqa
where
$\beta_d$
is a constant that is finite in $2 \leq d < 3$.
For other one-loop diagrams, say the one-loop fermion self-energy,
one should use \eq{babos} for the internal boson propagator.
Since the boson propagator itself depends on the coupling $g$,
the one-loop fermion self-energy becomes order of $g^{4/3}$ instead of $g^2$.
The non-analyticity in the coupling is a sign that 
some quantum effects for the boson self-energy are included non-perturbatively.
At the next order, one includes not only 
the two-loop fermion self-energy computed with the one-loop dressed boson propagator,
but also the correction generated from updating 
the boson propagator inside the one-loop fermion self-energy graph
with the two-loop boson self-energy.
This unusual order of including Feynman diagrams 
is forced upon us by the dynamical structure of the theory.
Although the expansion is not organized by the number of loops,
the procedure guarantees that every Feynman diagram is included
once and only once order by order in $\ep$.

Requiring that the physical observables are finite for $0 \leq \ep \leq 1/2$,
we add local counter terms that remove poles in $1/\ep$.
The bare theory that generates finite physical observables
is given by the sum of the classical action
and the counter terms.
From the condition that the bare theory is independent of the scale 
at which the low-energy observables are defined,
one obtains the beta function that describes
the flow of the renormalized coupling 
as a function of scale.
To the order of $O(\ep^2)$, the beta function becomes
\bqa
\frac{d g}{d l} & = & \frac{\epsilon}{2} g
- 0.02920 \left( \frac{3}{2} - \epsilon \right) \frac{g^{7/3}}{N}
+ 0.01073 \left( \frac{3}{2} - \epsilon \right)  \frac{g^{11/3}}{N^2}, 
\eqa
where $l$ is the logarithmic length scale. 
It exhibits a stable interacting fixed point at
$\frac{ g^{* 4/3} }{N} =  11.417 \epsilon + 55.498 \epsilon^2$.
Because the two-point functions are insensitive to the size of the Fermi surface,
the propagators obey the scaling forms,
\bqa
D(k)  =  \frac{1}{  k_y^{2(1-\tilde \eta_\phi) }}
~~ f \left( \frac{| {\bf K} |^{1/z} }{k_y^2} \right), 
~~~~~
G(k)  =  \frac{1}{ | \delta_k |^{1- \tilde \eta_\psi  } }
~~ g \left( \frac{ | {\bf  K} |^{1/z} }{\delta_k} \right).
\label{Gk}
\eqa
Here the dynamical critical exponent is related to the anomalous dimension of the boson 
through $z  = \frac{3-2 \tilde \eta_\phi}{3 - 2 \epsilon}$.
The exact relation is due to the Ward identity
which originates from the fact that
the boson couples to the $(d-1)$-th component of
a conserved current
associated with the unbroken $U(1)$ symmetry,
$\Psi \rightarrow e^{i \varphi} \Psi$\cite{PhysRevB.82.075127,PhysRevB.82.045121,PhysRevB.88.245106}.
To the two-loop order, the anomalous dimensions are given by
$\tilde \eta_\phi = 0$,
$\tilde \eta_\psi = 0.1508 \epsilon^2$.
It happens that the anomalous dimension of boson remains zero,
and $z=\frac{3}{2}$ in two dimensions
up to the three-loop order.
This is due to the facts that
$\tilde \eta_\phi=0$ is exact for the single-patch theory\cite{PhysRevB.90.045121},
and the presence of the antipodal points in the two-patch theory does not play an important role
up to the three-loop order\cite{PhysRevB.88.245106}.
However, there is no symmetry that protects the scaling dimension of the boson,
and a non-trivial anomalous dimension,
which causes a deviation of $z$ from $3/2$, 
is expected to arise at higher loops\cite{PhysRevB.92.041112}.

Unlike the correlation functions that are local in momentum space,
thermodynamic responses and transport properties 
are sensitive to the size of the Fermi surface.
As low-energy excitation across 
the entire Fermi surface contribute 
to the thermodynamic responses and transport,
they violate the hyperscaling\cite{PhysRevB.88.245106,PhysRevB.95.075127}. 
Due to the emergent locality in momentum,
the free energy is linearly proportional to the size of Fermi surface
which provides a cut-off for $k_y$.
Since the specific heat has dimension 
$[c_V] = z (d-2) + \frac{3}{2}$, 
it scales with temperature as
$c_V  \sim  \Lambda_y T^{(d-2)+\frac{1}{z}}$,
where $\Lambda_y$ is a scale associated with the size of Fermi surface
which accounts for the violation of hyperscaling 
by dimension $[k_y] = 1/2$.

Under the tree-level patch scaling, 
a short-range four-fermion interaction 
has the scaling dimension $(3/2-d)$.
Although it is irrelevant by power counting 
in two dimensions and above,
this itself does not exclude 
perturbative superconducting instability
for the following reasons.
First, the pairing susceptibility 
does not obey hyperscaling
because all fermions near Fermi surface 
contribute to pairing 
with zero center of mass momentum.
The violation of hyperscaling 
enhances the effective scaling dimension
of the four-fermion interaction to $(2-d)$
through the scale $\Lambda_y$ 
\cite{PhysRevB.92.035141}.
Second, the soft collective mode 
feeds the four fermion coupling
in the pairing channel
with small momentum transfer\cite{PhysRevD.59.094019,PhysRevB.91.115111,PhysRevB.92.205104}.
While the non-Fermi liquid remains stable against pairing near $d=5/2$, 
the density of state increases, 
and the pairing interaction becomes stronger as $d=2$ is approached.
As a result, the system becomes unstable against pairing 
below a certain critical dimension 
which lies between  $2$ and $5/2$\cite{PhysRevB.91.115111,PhysRevLett.114.097001,PhysRevB.94.115138,2016arXiv161201542L}.

\subsection{Non-perturbative approach}
\label{HotSurfacenonPerturb}

Although the perturbative approach discussed in the previous section 
provides some insight into non-Fermi liquids,
the strongly interacting theory in two dimensions 
is far from being understood in general. 
One of the exceptions is a chiral non-Fermi liquid,
where exact critical exponents can be extracted.

\begin{figure}[!ht]
     \centering
     \includegraphics[scale=0.4]{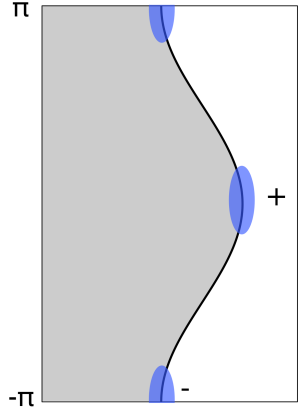}
     \caption{
A two-dimensional chiral Fermi surface
realized on the surface of a stack of quantum Hall layers. 
Generically, there are two points on the Fermi surface
which have a common tangent vector. 
For example, the points $+$ and $-$
have the same tangent vectors and they 
remain strongly coupled to each other  in the low energy
limit.
}
     \label{fig:chiral_FS}
\end{figure}

A chiral metal can arise
on the surface of three-dimensional
topological states. 
For example, the surface of a stack of  $\nu=1$ quantum Hall layers 
supports a two-dimensional metallic state
as the chiral edge modes acquire dispersion
in the direction perpendicular to the layers
through inter-layer tunnelings\cite{PhysRevLett.76.2782}.
The kinetic term for the chiral Fermi surface can be written as
\bqa
H_F = \sum_{K} \left( K_x - t \cos( K_y) \right) c_{j}^\dagger(\vec K) c_{j}(\vec K),
\label{ce}
\eqa
where $K_x$ ($K_y$) is the momentum along (perpendicular) to the edge,
$t$ is the nearest neighbor inter-layer tunneling,
and $j$ denotes an internal flavor such as orbital index.
In the presence of an internal symmetry,
the chiral metal can undergo a phase transition
associated with a spontaneous breaking of the symmetry,
and a chiral non-Fermi liquid arises at the quantum critical point\cite{PhysRevB.90.045121}.
The full low-energy theory is decomposed into
a set of decoupled patch theories.
Each patch theory describes
excitations near a set of points
on the Fermi surface
with a common tangent vector.
For the dispersion in \eq{ce}, there are generically two points 
with a parallel Fermi velocity
as is shown in \fig{fig:chiral_FS}. 
The theory that includes the patches 
centered at $\vec K_{F+}=(t,0)$ and $\vec K_{F-}=(-t,\pi)$
can be written as
\eqn{
S &=   
\int \frac{d^3 \vec{k}}{(2\pi)^3} ~ 
\lt(i \eta  k_0 +  k_x +   k_y^2 \rt) 
\psi_{js}^*(k)  \psi_{js}(k) \nn
& + \frac{1}{2} \int \frac{d^3 q}{(2\pi)^3} ~ q_y^2  ~|\phi_{\alpha}(q )|^2 \nn
& + g \int \frac{d^3 k d^3 q }{(2\pi)^6} ~
\phi_{\alpha}(q) ~
 \psi_{is}^*(k+ q) ~
 T^{\alpha}_{is;jt} ~   \psi_{jt}(k).
 \label{eq: b.f.action}
}
Here $\psi_{js}(k)$ is a chiral fermionic field with patch  $s=\pm$ 
which are related to the microscopic fermion field through
$\psi_{j+}(k_0,\vec k) = c_{j}(k_0,\vec K_{F+}+ \vec k)$,
$\psi_{j-}(k_0,\vec k) = c_{j}^*(-k_0,\vec K_{F-}- \vec k)$.
$\eta$ is a parameter that is introduced for a reason 
that will become clear soon.
$T^{\alpha}_{is;jt}$ is a matrix that describes 
the quantum number of the order parameter.
For the SU(2) internal symmetry,
$T^{\alpha}_{i+;j+} = \tau^\alpha_{ij}$,
 $T^{\alpha}_{i-;j-} = - (\tau^\alpha)^T_{ij}$
 and 
 $T^{\alpha}_{i+;j-} 
 =T^{\alpha}_{i-;j+}=0$,
 where $\tau^\alpha$ are the generators of SU(2) group with $\alpha=1,2,3$. 

In two dimensions,
the conventional perturbative expansion becomes unreliable 
as the coupling becomes large at low energies.
Since the interaction plays a dominant role,
we need to include the interaction
up front rather than treating it
as a perturbation to the kinetic energy. 
Therefore, we consider an interaction-driven scaling
in which the fermion-boson coupling is deemed marginal.
Under such a scaling, not all kinetic terms can be included as marginal operators.
At the one-loop order, 
the fermion self-energy is given by $i |k_0|^{2/3} \mbox{sign} (k_0)$,
which suggests that the $k_0$-linear term should be irrelevant.
The rest of the action is marginal 
if the dynamical critical exponent $z \equiv [k_0]$ 
and the dimensions of the fields are chosen to be
\bqa
z    =   \frac{3}{2},  ~~~ 
[\psi]  = [\phi]  =  - 2,
\label{eq:IS}
\eqa
with $[k_y] =1/2$ with $[k_x]=1$.
The scaling dimension of $k_y$ relative to $k_x$ is fixed by the sliding symmetry.

Under the interaction-driven scaling,
$\eta$ has scaling dimension $-1/2$,
and $\eta^{-2}$ plays the role of a UV cut-off
which is the only scale in \eq{eq: b.f.action}.
Since $\eta$ is irrelevant by power counting,
it is tempting to take the $\eta \rightarrow 0$ limit,
which is equivalent to taking the UV cut-off to infinity.
In the small $\eta$ limit, the theory has no scale.
However, this  alone neither guarantees the scale invariance of the theory
nor protects the scaling dimensions in 
\eq{eq:IS} 
from corrections
even if the theory remains scale invariant.
This is because of the scale anomaly.
Even though the classical action has no scale,
the quantum theory may not be well-defined without a UV cut-off.
If so, $\eta$ enters as a scale in the theory.
The scale can in principle generate a gap,
or modify the scaling dimensions 
in case the theory stays critical.

In the present theory,
it turns out that 
$\eta \rightarrow 0$ limit is well defined
even at the quantum level due to chirality.
First, the theory can not be gapped out
because the chiral gapless modes are protected
by the non-trivial topology (Chern number) in the bulk.
Second, the theory remains finite in the $\eta \rightarrow 0$ limit
because quantum fluctuations that involve particle-hole excitations 
have  limited phase space due to the chiral nature of the theory.
In particular, all internal frequencies in loops 
are bounded by external frequencies,
and loop integrations are finite in the small $\eta$ limit
due to a holomorphic structure 
inherited from the chirality\cite{PhysRevB.90.045121}.
This is analogous to the chiral Luttinger liquid in one dimension\cite{PhysRevB.41.12838}.
As a result, the theory remains critical,
and the quantum theory has no scale.
This implies that the  scaling dimensions in \eq{eq:IS} are exact.
The full fermion Green's function obeys the scaling form,
\bqa
G(k) = \frac{1}{\delta_k} ~ g \left(  \frac{ k_0^{2/3}}{ \delta_k } \right), \label{Gsc}
\eqa
where $\delta_k = k_x +  k_y^2$.
The universal function $g(x)$ is not fixed by the scaling, 
and the exact form of $g(x)$ can, in principle, be very different from 
what is inferred from the one-loop Green's function.
Nonetheless, the exponent is protected from quantum corrections.
This is a stable non-Fermi liquid,
which exhibits chaotic dynamics\cite{Patel21022017}.

\section{Theory of hot spots}

\begin{figure}
\begin{center}
\subfigure[]{\includegraphics[height=6cm,width=6cm]{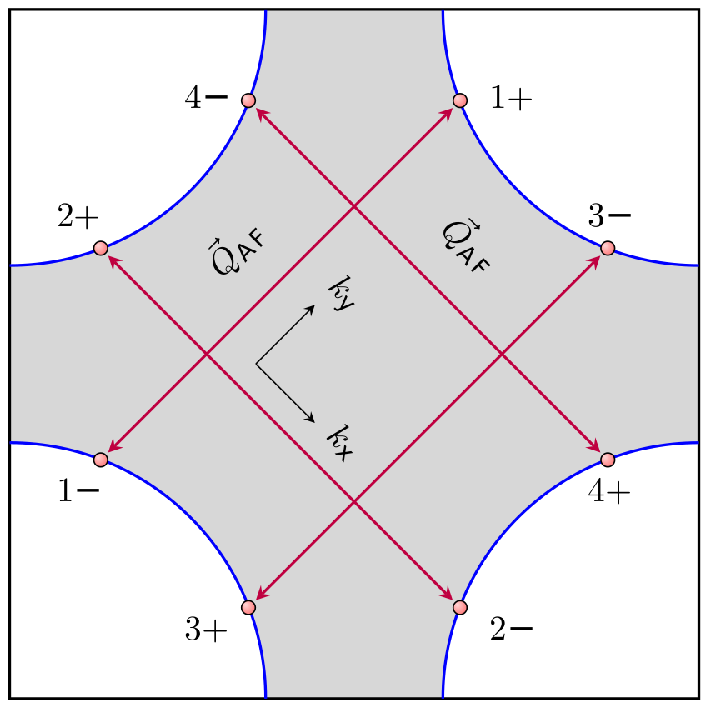} 
\label{fig:hot}
} ~~~~~
\subfigure[]{\includegraphics[height=3cm,width=3.5cm]{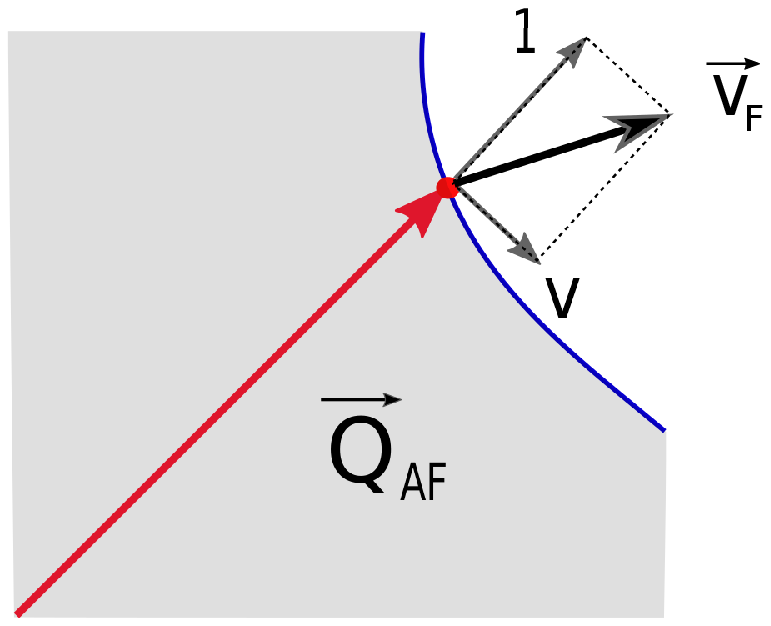} 
\label{fig:v}
}
\end{center}
\caption{
(a) Eight hot spots connected by 
the antiferromagnetic ordering vector $\vec Q_{AF}$ 
on a two-dimensional Fermi surface with $C_4$ symmetry.
(b) Decomposition of the Fermi velocity at hot spot $1+$
into the component parallel to $\vec Q_{AF}$
and the component perpendicular to $\vec Q_{AF}$.
By rescaling momentum, the component parallel to $\vec Q_{AF}$ is set to be one.
The component perpendicular to $\vec Q_{AF}$ denoted as $v$ measures 
the degree of local nesting between the hot spots 
connected by $\vec Q_{AF}$.
}
\label{fig:hot_spots}
\end{figure}

In this section, we turn to the second type of non-Fermi liquids,
where only a part of Fermi surface remains strongly coupled
with a gapless collective mode.
To be concrete, we consider the non-Fermi liquid
realized at the $SU(2)$ symmetric 
antiferromagnetic (AF) quantum critical point
with a commensurate wave vector in two dimensions.
With the $C_4$ symmetry, 
the low-energy degrees of freedom consist of 
the AF collective mode coupled to electrons near eight hot spots, 
which are the points on the Fermi surface connected by the AF ordering vector \cite{PhysRevLett.84.5608, PhysRevLett.93.255702, PhysRevB.82.075128, PhysRevB.91.125136}, 
as is shown in Fig. \ref{fig:hot}. 
The minimal model is written as
\beqa
\mathcal{S} = &&  
\sum_{n=1}^4 \sum_{m=\pm} \sum_{\sigma=\uparrow,\downarrow} 
\int \f{d^3k}{(2\pi)^3} ~ 
{\psi}^{(m)*}_{n,\s}(k)
\left[ ik_0 ~ + e^{m}_n(\vec k;v)  \right] 
\psi^{(m)}_{n,\sigma}(k)   \nn 
&&
+ \half \int \frac{d^3q}{(2\pi)^3} 
\left[ q_0^2 + c^2 | \vec q |^2 \right] \vec{\phi}(-q)  \cdot \vec{\phi}(q)   \nn 
&& + g_0 \sum_{n=1}^4 
\sum_{\s,\s'=\uparrow,\downarrow}
\int \f{d^3k}{(2\pi)^3} 
\f{d^3q}{(2\pi)^3} ~ 
\Bigl[
\vec{\phi}(q) \cdot 
{\psi}^{(+)*}_{n,\s}(k+q) \vec{\tau}_{\s,\s'}  
\psi^{(-)}_{n,\s'} (k) 
+ c.c. \Bigr] \nn 
&& + u_0
\int \f{d^3k}{(2\pi)^3}
\f{d^3p}{(2\pi)^3} 
\f{d^3q}{(2\pi)^3}
\lt[ \vec{\phi}(k+q) \cdot \vec{\phi}(p-q)\rt] 
\lt[ \vec{\phi}(-k) \cdot \vec{\phi}(-p) \rt]. 
\label{eq:3D_theory}
\eeqa
Here, $k = (k_0, \vec{k})$ denotes the Matsubara frequency and the two-dimensional momentum $\vec{k} = (k_x, k_y)$. 
$\psi_{n,\s}^{(m)}$ are the fermion fields that carry spin $\s = \uparrow, \downarrow$ at the hot spots labeled by $n=1,2,3,4, ~ m = \pm$. 
The coordinate axes have been chosen such that the ordering wave vector is 
$\vec{Q}_{AF} = \pm \sqrt{2} \pi \hat{k}_x, \pm \sqrt{2} \pi \hat{k}_y$ up to the reciprocal lattice vectors $\sqrt{2} \pi (\hat{k}_x \pm \hat{k}_y)$. 
With this choice the fermion dispersions are 
$e^{\pm}_1(\vec k;v) = -e^{\pm}_3(\vec k;v) = v k_x \pm k_y$, 
$e^{\pm}_2(\vec k;v) = -e^{\pm}_4(\vec k;v) = \mp k_x + v k_y$, 
where $\vec{k}$ is the momentum deviation from each hot spot. 
The Fermi velocity along the ordering vector has been set to be one by rescaling $\vec{k}$. 
$v$ is the component of Fermi velocity that is perpendicular to $\vec{Q}_{AF}$
as is shown in \fig{fig:v}. 
The curvature of the Fermi surface can be ignored, 
since the patches of Fermi surface connected by 
the ordering vector are not parallel to each other 
with $v \neq 0$.
$\vec{\phi}(q)$ is the boson field with three components of the  AF collective mode 
with frequency $q_0$ and momentum $\vec{Q}_{AF} + \vec{q}$. 
$\vec{\tau}$ represents the three generators of the $SU(2)$ group. 
$c$ is the velocity of the AF collective mode. 
$g_0$ is the Yukawa coupling between the collective mode and the electrons near the hot spots,
and  $u_0$ is the quartic coupling between the collective modes.

\subsection{Perturbative approach}
\label{HotSpotPerturb}

In order to access the interacting non-Fermi liquid perturbatively,
we deform the theory by tuning the number of co-dimensions 
of the Fermi surface\cite{PhysRevB.88.245106,PhysRevB.91.125136}.
Since the the curvature near the hot spots can be ignored,
the opposite sides of the Fermi surface
are locally nested with a translation by momentum $2 \vec K_F$.
This allows us to pair fermions on opposite sides of the Fermi surface into two component spinors,
 $\Psi_{1,\s} = (\psi_{1,\s}^{(+)}, \psi_{3,\s}^{(+)})^T$, 
 $\Psi_{2,\s} = (\psi_{2,\s}^{(+)}, \psi_{4,\s}^{(+)})^T$, 
 $\Psi_{3,\s} = (\psi_{1,\s}^{(-)}, - \psi_{3,\s}^{(-)})^T$, 
 $\Psi_{4,\s} = (\psi_{2,\s}^{(-)}, -\psi_{4,\s}^{(-)})^T$
 to cast the fermionic action of \eq{eq:3D_theory} to the spinor form,
 $S_F = \sum_{n=1}^4 \sum_{\sigma=\uparrow,\downarrow} \int \f{d^3k}{(2\pi)^3} ~ \bar{\Psi}_{n,\s}(k)
\left[ i \g_0 k_0 ~ + i \g_1 \eps_n(\vec{k};v)  \right] \Psi_{n,\sigma}(k)$, 
where $\g_0 = \s_y$ and $\g_1 = \s_x$,
$\bar{\Psi}_{n,\s} = \Psi^{\dagger}_{n,\s} \g_0$ 
with 
$\eps_1(\vec{k};v) = e_1^+(\vec{k};v)$, 
$\eps_2(\vec{k};v) = e_2^+(\vec{k};v)$, 
$\eps_3(\vec{k};v) = e_1^-(\vec{k};v)$, 
$\eps_4(\vec{k};v) = e_2^-(\vec{k};v)$. 
The spinor components are different from the ones used
in section \ref{HotSurfacenonPerturb}.
For hot Fermi surfaces, 
the local curvature of Fermi surface is important,
and the spinor is formed out of a particle and a hole 
to maintain the nesting in the dispersion 
of the two components in the spinor.
The theory in general dimensions can be written 
similarly as in \eq{act4}\cite{PhysRevB.88.245106,PhysRevB.91.125136}.
The upper critical dimension is three, 
and the perturbative non-Fermi liquid can be accessed
for small $\ep=(3-d)$.
The upper critical dimension is different from the one for the Ising-nematic critical metal
because the local curvature of Fermi surface is unimportant in the present case.

The action in general $d$ continuously interpolates 
the original AF critical metal in two dimensions 
to the AF critical semi-metal with line nodes in three dimensions.
The off-diagonal element of the kinetic term in general dimensions
corresponds to a $p$-wave charge density wave 
which gaps out the otherwise $(d-1)$-dimensional Fermi surface
into the line node. 
Similar to the Ising-nematic case, 
this breaks some internal symmetry.
The original theory in \eq{eq:3D_theory} has 
$SU(2) \times U(1)^4$ internal symmetry,
where $SU(2)$ is the spin rotation,
and the four $U(1)$'s refer to the separate conservations
of electron numbers in the four pairs 
of hot spots connected by $\vec Q_{AF}$.
On the other hand, the theory in $d>2$ 
has only $SU(2) \times U(1)^2$
as the off-diagonal kinetic term breaks two $U(1)$'s. 
One can ask how serious the symmetry breaking is 
in interpolating the results obtained in $d>2$ to $d=2$.
It turns out that the theory in any $2 \leq d < 3$ 
can be solved exactly
using a non-perturbative method 
to be discussed in the next section\cite{SDWgenD}.
The critical exponents obtained in general dimensions 
smoothly interpolate to the answer in $d=2$\cite{2016arXiv160806927S}.

\begin{figure}[h]
	\centering 
\includegraphics[width=2in]{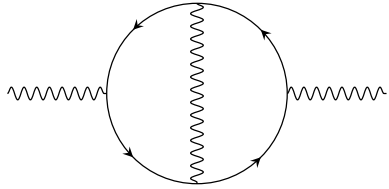}
	\caption{
Two-loop diagram that remains important
even to the leading order in $\ep$.
This diagram is nominally $\ep^2$ but becomes
enhanced to $\ep$ due to the emergent quasilocality
caused by the vanishing velocities.
}
	\label{fig:two-loop boson self energy}
\end{figure}
A systematic perturbative analysis has been done 
for this model\cite{PhysRevB.91.125136,2017arXiv170108218L}
and a related model with the $C_2$ symmetry\cite{PhysRevB.94.195135}.
Interestingly, all four parameters of the theory $\{ g,u,v,c \}$
flow to zero in the low energy limit.
In particular, the Fermi surface near the hot spots exhibits
an emergent nesting,
and the collective mode slows down 
due to the Landau damping.
Along with the emergent quasilocality,
the couplings also flow to zero 
such that $g^2/v \sim O(\ep)$, $u/c^2 = 0$ at the fixed point.
Unlike in relativistic field theories, 
it is not enough to consider only the one-loop 
graphs even to the leading order in $\ep$\cite{PhysRevB.91.125136,PhysRevB.94.195135,2017arXiv170108218L}.
This is because the vanishingly small velocities
can cause IR singularities at low energies
such that some higher-loop corrections remain important even to the leading order in $\ep$. 
It turns out that 
in the small $\ep$ limit
one needs to include a two-loop graph 
shown in  \fig{fig:two-loop boson self energy}
in addition to the one-loop graphs.
In the low energy limit,
the coupling and the velocities all flow to zero 
such that 
\beq
	\frac{g^2}{v} = 4 \pi \ep, 
	\quad \frac{g^2}{c^3} = \f{1}{16 \pi \, h_5^*}, 
	\quad \frac{v}{c} = 0
	\label{eq:fixed point values}
\eeq
with $h_5^* \approx 5.7 \times 10^{-4}$.
The two-loop diagram which is nominally order of $\ep^2$ 
becomes enhanced to the order of $\ep$ due to the quasilocality.
Other higher-loop graphs remain small in the small $\ep$ limit,
and the $\ep$-expansion is controlled
after the two-loop effect is included\cite{2017arXiv170108218L}.
At the fixed point,
the dynamical critical exponent 
and the scaling dimensions of the fundamental fields are given by
\bqa
z=1, 
~~~[\psi(k)]=-\frac{5-\ep}{2}, ~~~ [\phi(k)]=-(3-\ep).
\label{epscaling}
\eqa
It is noted that the fermion retains its classical dimension
while the boson acquires a non-trivial anomalous dimension
$\eta_\phi=\ep/2$.
The boson is heavily dressed by 
particle-hole excitations, 
and it no longer propagates as a coherent excitation.
On the other hand, the electrons remain largely intact
in the low energy limit.

The most striking outcome revealed by the controlled expansion 
is the emergent hierarchy among three velocities :
the component of the Fermi velocity along $\vec Q_{AF}$ which is set to be one,
the velocity of the collective mode ($c$), 
and the component of the Fermi velocity perpendicular to $\vec Q_{AF}$ ($v$).
In the low energy limit, $c$ and $v$ flow to zero such that $v \ll c \ll 1$.
This poses both challenge and opportunity.
On the one hand, higher-loop diagrams are not trivially suppressed
even to the leading order  in $\ep$
due to the infrared singularity caused by the emergent locality.
On the other hand, the ratios among velocities
can be used as small parameters to solve
the strongly interacting theory
even in two dimensions\cite{PhysRevB.78.064512}.

\subsection{Non-perturbative approach}
\label{HotSpotnonPerturb}

In order to tackle the theory in two dimensions directly,
we start with an interaction-driven scaling 
that incorporates the interaction up front.
Once the fermion-boson coupling is deemed marginal,
one cannot keep all the kinetic terms as marginal operators.
In choosing the marginal terms in the kinetic energy, 
we make a choice different from \ref{HotSurfacenonPerturb}.
Here we choose a scaling in which 
all the fermion kinetic term is kept marginal 
at the expense of discarding the entire boson kinetic term as irrelevant terms. 
This choice is inspired from \eq{epscaling}.
The marginality of the fermion kinetic term and the fermion-boson coupling 
uniquely fixes the dynamical critical exponent and the scaling dimensions,
\bqa
z=1, ~~~~ [\psi(k)] = [\phi(k)] = -2.
\label{AN1}
\eqa
This Ansatz is obtained by setting $\ep=1$ in \eq{epscaling}.
In general, one would expect higher-order corrections in $\ep$.
In the present case, it turns out that \eq{AN1} becomes exact 
in the low-energy limit
because of the emergent hierarchy among the velocities\cite{2016arXiv160806927S}.

Under \eq{AN1}, the entire boson kinetic term and the quartic coupling are irrelevant.
The minimal action which includes only marginal terms is written as
\begin{align}
  \mathcal{S} &= 
\sum_{n=1}^4 
\sum_{\sigma=\uparrow, \downarrow}
\int 
d k~
\bar{\Psi}_{n,\sigma}(k) 
\Bigl[ i \gamma_0 k_0 + i \gamma_{1} \varepsilon_n(\vec{k}) \Bigr] 
\Psi_{n,\sigma}(k)
\nn 
  & + i  \sqrt{ \frac{\pi v}{2} }  
  \sum_{n=1}^4
\sum_{\sigma,\sigma'}
\int 
d k d q ~
\bar{\Psi}_{\bar n,\sigma}(k+q) 
\vec \tau_{\sigma,\sigma'}
\gamma_{1} 
\Psi_{n,\sigma'} (k) 
\cdot 
\vec \phi(q).
\label{eq:min_theory}
\end{align}
Here, we rescale the boson field to set the fermion-boson coupling to be proportional to $\sqrt{v}$.
This is a convenient choice because
the interaction is screened
such that $g^2$ becomes $O(v)$ 
in the low-energy limit\cite{PhysRevB.91.125136}.
Although one can tune $g$ and $v$ independently in a microscopic theory,
they flow to a universal line defined by $g^2 \sim v$ at low energies\cite{2017arXiv170108218L}.

\begin{figure}[!ht]
\centering
\subfigure[]{
 \includegraphics[scale=1.2]{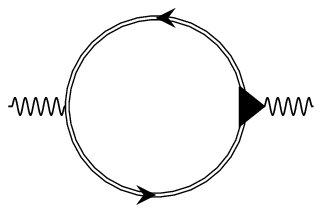}
 \label{fig:SD}
 }~~~~~~
 \subfigure[]{
 \includegraphics[scale=1.2]{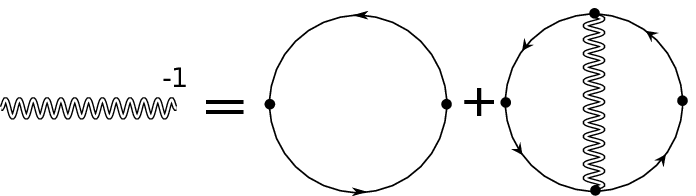}
 \label{fig:SD2}
}
 \caption{
(a) The exact boson self-energy.
The double line is the fully dressed fermion propagator.
The triangle represents the fully dressed vertex.
(b) The reduced Schwinger-Dyson equation 
that replaces (a) in the small $v/c$ limit. 
}
\end{figure}

In the absence of the bare action for boson, 
dynamics for the collective mode is entirely generated from particle-hole excitations
as is shown in \fig{fig:SD}.
The exact Schwinger-Dyson equation for the boson propagator reads 
\bqa
D(q)^{-1} &=& 
m_{CT}
-  \pi v \sum_n
\int dk ~ 
\mbox{Tr}\lt[ \gamma_{1} G_{\bar n}(k + q) 
\Gamma(k,q)  G_n(k) \rt] .
\label{eq:SD}
\eqa
Here $D(k)$, $G(k)$ and $\Gamma(k,q)$ represent 
the fully dressed propagators of the boson and the fermion,
and the vertex function, respectively.
$m_{CT}$ is a mass counter term that is added 
to tune the renormalized mass to zero. 
The trace is over the spinor indices.
An Ansatz for $D(q)$ that is consistent with \eq{AN1} is
\bqa
D(q)^{-1} = |q_0| + c(v) \Big[ |q_x| + |q_y| \Big].
\label{eq:D}
\eqa
One needs to show that \eq{eq:D} indeed satisfies the Schwinger-Dyson equation,
and determine the `velocity' of the strongly damped collective mode 
$c(v)$ as a function of $v$.
Although the Schwinger-Dyson equation cannot be solved in general,
it can be solved in the limit $v/c(v)$ is small.
Given that we don't know whether $v/c(v)$ flows to zero in two dimensions yet,
we first solve the Schwinger-Dyson equation 
under the assumption that  $v \ll c(v) \ll 1$,
and then prove that $v$, $c(v)$ and $v/c(v)$ indeed flow to zero using the 
solution of the Schwinger-Dyson equation.

In the small $v/c(v)$ limit,
the Schwinger-Dyson equation
can be simplified.
In particular, the full fermion propagator
and the vertex function in \eq{eq:SD} can be replaced with
the bare fermion propagator
and the one-loop dressed vertex function respectively
to the leading order in $v/c(v)$.
This can be understood
in the following way\cite{2017arXiv170108218L}.
The interaction mixes the fermions in patches connected by $\vec Q_{AF}$.
This renormalizes the velocity of fermions 
such that the Fermi surface becomes locally nested near the hot spots\cite{
PhysRevLett.84.5608,
doi:10.1080/0001873021000057123,
PhysRevB.82.075128,
PhysRevB.93.165114,
PhysRevB.91.125136,
2017arXiv170108218L}.   
A well-nested Fermi surface provides a large phase space for soft particle-hole excitations
that screen the interaction to $g^2 \sim v$.
The flat spectrum of the particle-hole excitations also slows down the collective mode.
The slow velocity of the collective mode 
enhances the feedback of the collective mode to the fermion.
However, it is not strong enough to overcome the suppression caused by large screening. 
This is because the velocity of the collective mode remains larger than $v$
due to the anti-screening effect of the vertex correction 
in \fig{fig:two-loop boson self energy} that
speeds up the collective mode\cite{PhysRevB.94.195135,2016arXiv160806927S}.
Therefore, the fermion self-energy and the vertex correction
is suppressed by $g^2/c(v) \sim v/c(v)$.
Although both fermion self-energy and vertex corrections are small in the small $v/c(v)$ limit, 
the leading vertex correction needs to be kept in \eq{eq:SD}.
This is because the one-loop boson self-energy with the bare vertex 
is independent of momentum. 
In order to keep the minimal momentum dependence in the boson propagator,
the one-loop vertex correction should be included in the Schwinger-Dyson equation.
Therefore, the full Dyson equation is reduced
to \fig{fig:SD2} in the small $v/c(v)$ limit.

The solution to the reduced Schwinger-Dyson equation is indeed given by \eq{eq:D}
with
\bqa
c(v) = \frac{1}{4} \sqrt{  v \log (1/v) },
\label{eq:cv}
\eqa
and the hierarchy is satisfied, $v \ll c(v) \ll 1$
in the small $v$ limit.
The remaining job is to show that $v$ flows to zero.
The RG flow of $v$ is determined  
from the fermion self-energy 
and the vertex correction
computed with the dressed boson propagator. 
Under the RG flow, $v$ flows to zero
with increasing logarithmic length scale $l$ as 
\bqa
\frac{dv}{dl} =  -\f{6}{\pi^2} \, v^{2} \log\lt( \f1{c(v)} \rt).
\label{eq:beta2}
\eqa
This completes the cycle of self-consistency.
\eq{eq:D} obtained in the small $v$ limit becomes asymptotically 
exact in the low-energy limit
within a nonzero basin of attraction in the space of $v$ near $v=0$.
We emphasize that the theory is still strongly coupled in
the  small $v$ limit in that the conventional perturbative expansion
is controlled by $g^2/v \sim 1$.
One manifestation of this is that  \fig{fig:SD2}
includes an infinite series of diagrams 
that are all same order of magnitude.

If the bare value of $v$ is small, 
there exist well-separated energy scales
which dictate multiple crossovers.
The first crossover is set by 
the competition between 
\eq{eq:D} and the irrelevant local kinetic term 
$\frac{|\vec q|^2}{\tilde \Lambda}$,
where $\tilde \Lambda$ represents an energy scale 
introduced by rescaling the boson field 
from \eq{eq:3D_theory}.
For $\omega < E_b^*$ with $E_b^* = c^2 \tilde \Lambda$,
the terms linear in frequency and momentum dominate.
The second energy scale is the superconducting transition temperature.
The spin fluctuations renormalize pairing interactions near the hot spots, and  
enhance the $d$-wave superconductivity\cite{scalapino1986d,PhysRevB.34.6554,doi:10.1143/JPSJ.59.2905,Berg21122012,2015arXiv151204541L,PhysRevLett.117.097002}. 
Due the gapless collective mode which  mediates a marginal interaction,
the pairing vertex  at frequency $\omega$ is enhanced by double logarithms,
$\alpha \frac{v}{c} \log \frac{\Lambda}{\omega} \log \frac{ E_b^* }{\omega}$ 
with $\alpha \sim 1$\cite{PhysRevD.59.094019,PhysRevB.46.14803,PhysRevB.72.174520}.
This gives $T_c \sim  c \sqrt{ \Lambda  \tilde \Lambda} 
e^{-   \frac{1}{2\sqrt{\alpha}} 
\frac{\log^{1/2}1/v}{ v^{1/4}}  }$.
The third energy scale, denoted as $E_f^*$, 
is the one below which 
the behavior of the fermions at the hot spots 
deviates from the Fermi liquid one.
For a small $v$, 
the leading order self-energy correction to the fermion propagator
is $\frac{3}{4\pi} \frac{v}{c} \omega \log \frac{\Lambda}{\omega}$,
which becomes larger than the bare term
for $\omega < E_f^*$ with 
$E_f^* \sim \Lambda e^{-\frac{\pi}{3}\sqrt{ \frac{\log1/v}{v} }}$.
This scale is small 
because the fermion is only weakly
perturbed by the collective mode.
Finally, the value of $v$ changes appreciably
below $E_v^* \sim \Lambda e^{-\frac{1}{v \log 1/v}}$,
which is determined from \eq{eq:beta2}.

In the small $v$ limit, 
there is a hierarchy among the energy scales,
$E_v^* \ll E_f^* \ll T_c \ll E_b^* $.
This suggests that 
the system undergoes a superconducting transition
before the fermions at the hot spots lose coherence
or $v$ changes appreciably.
On the other hand, 
there is a large window between $T_c$ and $E_b^*$
within which the universal scaling 
for the collective mode given by \eq{eq:D} is obeyed.
The size of the energy window for the critical scaling 
is non-universal due to the slow flow of $v$, 
and it depends on the bare value of $v$.
It is expected that 
the energy window for the $z=1$ critical scaling above $T_c$ is larger
in materials whose bare Fermi surfaces 
are closer to  perfect nesting near the hot spots.

\section{Conclusion}

Low-energy effective theories for non-Fermi liquids
have unusual features that originate
from the presence of infinitely many gapless modes.
Strong infrared quantum fluctuations 
modify the way in which quantum corrections are organized 
compared to relativistic quantum field theories.
Quantum corrections that are normally considered to be higher-order effects
should be included to the leading order even near the upper critical dimensions.
Interestingly, this strong quantum effect 
is responsible for a remarkable simplicity that emerges
in the low energy limit.
In particular, the interaction-driven scaling 
which takes into account interactions ahead of the kinetic term
at the `tree'-level
gives exact critical exponents
in some strongly coupled non-Fermi liquids.

It appears that
the solvability of the strongly coupled field theories 
is rooted to the fact that the interaction-driven scaling 
is as far as interactions can do in critical states.
The low-energy scaling is the result of
competition between 
the interactions that tend to pin particles in real space
and the kinetic energy that promotes delocalization.
When the kinetic energy dominates as in Fermi liquids,
the interactions play only limited roles,
and the scaling that prioritizes the kinetic energy emerges.
If interactions are strong,
one can have a localized state 
such as Mott insulators depending on the density of particles.
However, if an insulating state is not an option 
for itinerant quantum critical points,
the way particles propagate is conformed by the dominant interaction 
so that the bare kinetic term can be completely ignored,
which gives rise to the interaction-driven scaling.
It remains to be understood 
how large the class of strongly coupled theories
that obey the same principle is.
It will be also interesting to search for 
yet another simple dynamical principles
that may open up new windows 
to non-perturbative physics 
of strongly interacting systems.
On the phenomenological side,
many open questions are waiting to be addressed.
For example, 
non-equilibrium phenomena in general,
and the interplay of non-Fermi liquid behavior
with impurities
are interesting open problems.

{\bf Acknowledgment}
The author thanks
Denis Dalidovich,
Peter Lunts,
Ipsita Mandal,
Andres Schlief
and 
Shouvik Sur,
for collaborations on the subject.
The research was supported by 
the Natural Sciences and Engineering Research Council of 
Canada.
Research at the Perimeter Institute is supported 
in part by the Government of Canada 
through Industry Canada, 
and by the Province of Ontario through the
Ministry of Research and Information.

\bibliography{NFL}

\begin{thebibliography}{65}%
\makeatletter
\providecommand \@ifxundefined [1]{%
 \@ifx{#1\undefined}
}%
\providecommand \@ifnum [1]{%
 \ifnum #1\expandafter \@firstoftwo
 \else \expandafter \@secondoftwo
 \fi
}%
\providecommand \@ifx [1]{%
 \ifx #1\expandafter \@firstoftwo
 \else \expandafter \@secondoftwo
 \fi
}%
\providecommand \natexlab [1]{#1}%
\providecommand \enquote  [1]{``#1''}%
\providecommand \bibnamefont  [1]{#1}%
\providecommand \bibfnamefont [1]{#1}%
\providecommand \citenamefont [1]{#1}%
\providecommand \href@noop [0]{\@secondoftwo}%
\providecommand \href [0]{\begingroup \@sanitize@url \@href}%
\providecommand \@href[1]{\@@startlink{#1}\@@href}%
\providecommand \@@href[1]{\endgroup#1\@@endlink}%
\providecommand \@sanitize@url [0]{\catcode `\\12\catcode `\$12\catcode
  `\&12\catcode `\#12\catcode `\^12\catcode `\_12\catcode `\%12\relax}%
\providecommand \@@startlink[1]{}%
\providecommand \@@endlink[0]{}%
\providecommand \url  [0]{\begingroup\@sanitize@url \@url }%
\providecommand \@url [1]{\endgroup\@href {#1}{\urlprefix }}%
\providecommand \urlprefix  [0]{URL }%
\providecommand \Eprint [0]{\href }%
\providecommand \doibase [0]{http://dx.doi.org/}%
\providecommand \selectlanguage [0]{\@gobble}%
\providecommand \bibinfo  [0]{\@secondoftwo}%
\providecommand \bibfield  [0]{\@secondoftwo}%
\providecommand \translation [1]{[#1]}%
\providecommand \BibitemOpen [0]{}%
\providecommand \bibitemStop [0]{}%
\providecommand \bibitemNoStop [0]{.\EOS\space}%
\providecommand \EOS [0]{\spacefactor3000\relax}%
\providecommand \BibitemShut  [1]{\csname bibitem#1\endcsname}%
\let\auto@bib@innerbib\@empty
\bibitem [{\citenamefont {Landau}(1957)}]{LFL}%
  \BibitemOpen
  \bibfield  {author} {\bibinfo {author} {\bibfnamefont {L.}~\bibnamefont
  {Landau}},\ }\href@noop {} {\bibfield  {journal} {\bibinfo  {journal} {Sov.
  Phys. JETP}\ }\textbf {\bibinfo {volume} {3}},\ \bibinfo {pages} {920}
  (\bibinfo {year} {1957})}\BibitemShut {NoStop}%
\bibitem [{\citenamefont {{Polchinski}}(1992)}]{1992hep.th...10046P}%
  \BibitemOpen
  \bibfield  {author} {\bibinfo {author} {\bibfnamefont {J.}~\bibnamefont
  {{Polchinski}}},\ }\href@noop {} {\bibfield  {journal} {\bibinfo  {journal}
  {ArXiv High Energy Physics - Theory e-prints}\ } (\bibinfo {year} {1992})},\
  \Eprint {http://arxiv.org/abs/hep-th/9210046} {hep-th/9210046} \BibitemShut
  {NoStop}%
\bibitem [{\citenamefont {Shankar}(1994)}]{RevModPhys.66.129}%
  \BibitemOpen
  \bibfield  {author} {\bibinfo {author} {\bibfnamefont {R.}~\bibnamefont
  {Shankar}},\ }\href {\doibase 10.1103/RevModPhys.66.129} {\bibfield
  {journal} {\bibinfo  {journal} {Rev. Mod. Phys.}\ }\textbf {\bibinfo {volume}
  {66}},\ \bibinfo {pages} {129} (\bibinfo {year} {1994})}\BibitemShut
  {NoStop}%
\bibitem [{\citenamefont {Hertz}(1976)}]{PhysRevB.14.1165}%
  \BibitemOpen
  \bibfield  {author} {\bibinfo {author} {\bibfnamefont {J.~A.}\ \bibnamefont
  {Hertz}},\ }\href {\doibase 10.1103/PhysRevB.14.1165} {\bibfield  {journal}
  {\bibinfo  {journal} {Phys. Rev. B}\ }\textbf {\bibinfo {volume} {14}},\
  \bibinfo {pages} {1165} (\bibinfo {year} {1976})}\BibitemShut {NoStop}%
\bibitem [{\citenamefont {Millis}(1993)}]{PhysRevB.48.7183}%
  \BibitemOpen
  \bibfield  {author} {\bibinfo {author} {\bibfnamefont {A.~J.}\ \bibnamefont
  {Millis}},\ }\href {\doibase 10.1103/PhysRevB.48.7183} {\bibfield  {journal}
  {\bibinfo  {journal} {Phys. Rev. B}\ }\textbf {\bibinfo {volume} {48}},\
  \bibinfo {pages} {7183} (\bibinfo {year} {1993})}\BibitemShut {NoStop}%
\bibitem [{\citenamefont {L\"ohneysen}\ \emph {et~al.}(2007)\citenamefont
  {L\"ohneysen}, \citenamefont {Rosch}, \citenamefont {Vojta},\ and\
  \citenamefont {W\"olfle}}]{RevModPhys.79.1015}%
  \BibitemOpen
  \bibfield  {author} {\bibinfo {author} {\bibfnamefont {H.~v.}\ \bibnamefont
  {L\"ohneysen}}, \bibinfo {author} {\bibfnamefont {A.}~\bibnamefont {Rosch}},
  \bibinfo {author} {\bibfnamefont {M.}~\bibnamefont {Vojta}}, \ and\ \bibinfo
  {author} {\bibfnamefont {P.}~\bibnamefont {W\"olfle}},\ }\href {\doibase
  10.1103/RevModPhys.79.1015} {\bibfield  {journal} {\bibinfo  {journal} {Rev.
  Mod. Phys.}\ }\textbf {\bibinfo {volume} {79}},\ \bibinfo {pages} {1015}
  (\bibinfo {year} {2007})}\BibitemShut {NoStop}%
\bibitem [{\citenamefont {Stewart}(2001)}]{RevModPhys.73.797}%
  \BibitemOpen
  \bibfield  {author} {\bibinfo {author} {\bibfnamefont {G.~R.}\ \bibnamefont
  {Stewart}},\ }\href {\doibase 10.1103/RevModPhys.73.797} {\bibfield
  {journal} {\bibinfo  {journal} {Rev. Mod. Phys.}\ }\textbf {\bibinfo {volume}
  {73}},\ \bibinfo {pages} {797} (\bibinfo {year} {2001})}\BibitemShut
  {NoStop}%
\bibitem [{\citenamefont {Senthil}(2008)}]{PhysRevB.78.035103}%
  \BibitemOpen
  \bibfield  {author} {\bibinfo {author} {\bibfnamefont {T.}~\bibnamefont
  {Senthil}},\ }\href {\doibase 10.1103/PhysRevB.78.035103} {\bibfield
  {journal} {\bibinfo  {journal} {Phys. Rev. B}\ }\textbf {\bibinfo {volume}
  {78}},\ \bibinfo {pages} {035103} (\bibinfo {year} {2008})}\BibitemShut
  {NoStop}%
\bibitem [{\citenamefont {Holstein}\ \emph {et~al.}(1973)\citenamefont
  {Holstein}, \citenamefont {Norton},\ and\ \citenamefont
  {Pincus}}]{PhysRevB.8.2649}%
  \BibitemOpen
  \bibfield  {author} {\bibinfo {author} {\bibfnamefont {T.}~\bibnamefont
  {Holstein}}, \bibinfo {author} {\bibfnamefont {R.~E.}\ \bibnamefont
  {Norton}}, \ and\ \bibinfo {author} {\bibfnamefont {P.}~\bibnamefont
  {Pincus}},\ }\href {\doibase 10.1103/PhysRevB.8.2649} {\bibfield  {journal}
  {\bibinfo  {journal} {Phys. Rev. B}\ }\textbf {\bibinfo {volume} {8}},\
  \bibinfo {pages} {2649} (\bibinfo {year} {1973})}\BibitemShut {NoStop}%
\bibitem [{\citenamefont {Reizer}(1989)}]{PhysRevB.40.11571}%
  \BibitemOpen
  \bibfield  {author} {\bibinfo {author} {\bibfnamefont {M.~Y.}\ \bibnamefont
  {Reizer}},\ }\href {\doibase 10.1103/PhysRevB.40.11571} {\bibfield  {journal}
  {\bibinfo  {journal} {Phys. Rev. B}\ }\textbf {\bibinfo {volume} {40}},\
  \bibinfo {pages} {11571} (\bibinfo {year} {1989})}\BibitemShut {NoStop}%
\bibitem [{\citenamefont {Altshuler}\ \emph {et~al.}(1994)\citenamefont
  {Altshuler}, \citenamefont {Ioffe},\ and\ \citenamefont
  {Millis}}]{PhysRevB.50.14048}%
  \BibitemOpen
  \bibfield  {author} {\bibinfo {author} {\bibfnamefont {B.~L.}\ \bibnamefont
  {Altshuler}}, \bibinfo {author} {\bibfnamefont {L.~B.}\ \bibnamefont
  {Ioffe}}, \ and\ \bibinfo {author} {\bibfnamefont {A.~J.}\ \bibnamefont
  {Millis}},\ }\href {\doibase 10.1103/PhysRevB.50.14048} {\bibfield  {journal}
  {\bibinfo  {journal} {Phys. Rev. B}\ }\textbf {\bibinfo {volume} {50}},\
  \bibinfo {pages} {14048} (\bibinfo {year} {1994})}\BibitemShut {NoStop}%
\bibitem [{\citenamefont {Kim}\ \emph {et~al.}(1994)\citenamefont {Kim},
  \citenamefont {Furusaki}, \citenamefont {Wen},\ and\ \citenamefont
  {Lee}}]{PhysRevB.50.17917}%
  \BibitemOpen
  \bibfield  {author} {\bibinfo {author} {\bibfnamefont {Y.~B.}\ \bibnamefont
  {Kim}}, \bibinfo {author} {\bibfnamefont {A.}~\bibnamefont {Furusaki}},
  \bibinfo {author} {\bibfnamefont {X.-G.}\ \bibnamefont {Wen}}, \ and\
  \bibinfo {author} {\bibfnamefont {P.~A.}\ \bibnamefont {Lee}},\ }\href
  {\doibase 10.1103/PhysRevB.50.17917} {\bibfield  {journal} {\bibinfo
  {journal} {Phys. Rev. B}\ }\textbf {\bibinfo {volume} {50}},\ \bibinfo
  {pages} {17917} (\bibinfo {year} {1994})}\BibitemShut {NoStop}%
\bibitem [{\citenamefont {Lee}(1989)}]{PhysRevLett.63.680}%
  \BibitemOpen
  \bibfield  {author} {\bibinfo {author} {\bibfnamefont {P.~A.}\ \bibnamefont
  {Lee}},\ }\href {\doibase 10.1103/PhysRevLett.63.680} {\bibfield  {journal}
  {\bibinfo  {journal} {Phys. Rev. Lett.}\ }\textbf {\bibinfo {volume} {63}},\
  \bibinfo {pages} {680} (\bibinfo {year} {1989})}\BibitemShut {NoStop}%
\bibitem [{\citenamefont {Polchinski}(1994)}]{polchinski1994low}%
  \BibitemOpen
  \bibfield  {author} {\bibinfo {author} {\bibfnamefont {J.}~\bibnamefont
  {Polchinski}},\ }\href@noop {} {\bibfield  {journal} {\bibinfo  {journal}
  {Nuclear Physics B}\ }\textbf {\bibinfo {volume} {422}},\ \bibinfo {pages}
  {617} (\bibinfo {year} {1994})}\BibitemShut {NoStop}%
\bibitem [{\citenamefont {Lee}\ and\ \citenamefont
  {Nagaosa}(1992)}]{PhysRevB.46.5621}%
  \BibitemOpen
  \bibfield  {author} {\bibinfo {author} {\bibfnamefont {P.~A.}\ \bibnamefont
  {Lee}}\ and\ \bibinfo {author} {\bibfnamefont {N.}~\bibnamefont {Nagaosa}},\
  }\href {\doibase 10.1103/PhysRevB.46.5621} {\bibfield  {journal} {\bibinfo
  {journal} {Phys. Rev. B}\ }\textbf {\bibinfo {volume} {46}},\ \bibinfo
  {pages} {5621} (\bibinfo {year} {1992})}\BibitemShut {NoStop}%
\bibitem [{\citenamefont {Nayak}\ and\ \citenamefont
  {Wilczek}(1994)}]{nayak1994non}%
  \BibitemOpen
  \bibfield  {author} {\bibinfo {author} {\bibfnamefont {C.}~\bibnamefont
  {Nayak}}\ and\ \bibinfo {author} {\bibfnamefont {F.}~\bibnamefont
  {Wilczek}},\ }\href@noop {} {\bibfield  {journal} {\bibinfo  {journal}
  {Nuclear Physics B}\ }\textbf {\bibinfo {volume} {417}},\ \bibinfo {pages}
  {359} (\bibinfo {year} {1994})}\BibitemShut {NoStop}%
\bibitem [{\citenamefont {Lee}(2009)}]{PhysRevB.80.165102}%
  \BibitemOpen
  \bibfield  {author} {\bibinfo {author} {\bibfnamefont {S.-S.}\ \bibnamefont
  {Lee}},\ }\href {\doibase 10.1103/PhysRevB.80.165102} {\bibfield  {journal}
  {\bibinfo  {journal} {Phys. Rev. B}\ }\textbf {\bibinfo {volume} {80}},\
  \bibinfo {pages} {165102} (\bibinfo {year} {2009})}\BibitemShut {NoStop}%
\bibitem [{\citenamefont {Metlitski}\ and\ \citenamefont
  {Sachdev}(2010{\natexlab{a}})}]{PhysRevB.82.075127}%
  \BibitemOpen
  \bibfield  {author} {\bibinfo {author} {\bibfnamefont {M.~A.}\ \bibnamefont
  {Metlitski}}\ and\ \bibinfo {author} {\bibfnamefont {S.}~\bibnamefont
  {Sachdev}},\ }\href {\doibase 10.1103/PhysRevB.82.075127} {\bibfield
  {journal} {\bibinfo  {journal} {Phys. Rev. B}\ }\textbf {\bibinfo {volume}
  {82}},\ \bibinfo {pages} {075127} (\bibinfo {year}
  {2010}{\natexlab{a}})}\BibitemShut {NoStop}%
\bibitem [{\citenamefont {Mross}\ \emph {et~al.}(2010)\citenamefont {Mross},
  \citenamefont {McGreevy}, \citenamefont {Liu},\ and\ \citenamefont
  {Senthil}}]{PhysRevB.82.045121}%
  \BibitemOpen
  \bibfield  {author} {\bibinfo {author} {\bibfnamefont {D.~F.}\ \bibnamefont
  {Mross}}, \bibinfo {author} {\bibfnamefont {J.}~\bibnamefont {McGreevy}},
  \bibinfo {author} {\bibfnamefont {H.}~\bibnamefont {Liu}}, \ and\ \bibinfo
  {author} {\bibfnamefont {T.}~\bibnamefont {Senthil}},\ }\href {\doibase
  10.1103/PhysRevB.82.045121} {\bibfield  {journal} {\bibinfo  {journal} {Phys.
  Rev. B}\ }\textbf {\bibinfo {volume} {82}},\ \bibinfo {pages} {045121}
  (\bibinfo {year} {2010})}\BibitemShut {NoStop}%
\bibitem [{\citenamefont {Jiang}\ \emph {et~al.}(2013)\citenamefont {Jiang},
  \citenamefont {Block}, \citenamefont {Mishmash}, \citenamefont {Garrison},
  \citenamefont {Sheng}, \citenamefont {Motrunich},\ and\ \citenamefont
  {Fisher}}]{jiang2013non}%
  \BibitemOpen
  \bibfield  {author} {\bibinfo {author} {\bibfnamefont {H.-C.}\ \bibnamefont
  {Jiang}}, \bibinfo {author} {\bibfnamefont {M.~S.}\ \bibnamefont {Block}},
  \bibinfo {author} {\bibfnamefont {R.~V.}\ \bibnamefont {Mishmash}}, \bibinfo
  {author} {\bibfnamefont {J.~R.}\ \bibnamefont {Garrison}}, \bibinfo {author}
  {\bibfnamefont {D.}~\bibnamefont {Sheng}}, \bibinfo {author} {\bibfnamefont
  {O.~I.}\ \bibnamefont {Motrunich}}, \ and\ \bibinfo {author} {\bibfnamefont
  {M.~P.}\ \bibnamefont {Fisher}},\ }\href@noop {} {\bibfield  {journal}
  {\bibinfo  {journal} {Nature}\ }\textbf {\bibinfo {volume} {493}},\ \bibinfo
  {pages} {39} (\bibinfo {year} {2013})}\BibitemShut {NoStop}%
\bibitem [{\citenamefont {Dalidovich}\ and\ \citenamefont
  {Lee}(2013)}]{PhysRevB.88.245106}%
  \BibitemOpen
  \bibfield  {author} {\bibinfo {author} {\bibfnamefont {D.}~\bibnamefont
  {Dalidovich}}\ and\ \bibinfo {author} {\bibfnamefont {S.-S.}\ \bibnamefont
  {Lee}},\ }\href {\doibase 10.1103/PhysRevB.88.245106} {\bibfield  {journal}
  {\bibinfo  {journal} {Phys. Rev. B}\ }\textbf {\bibinfo {volume} {88}},\
  \bibinfo {pages} {245106} (\bibinfo {year} {2013})}\BibitemShut {NoStop}%
\bibitem [{\citenamefont {Sur}\ and\ \citenamefont
  {Lee}(2014)}]{PhysRevB.90.045121}%
  \BibitemOpen
  \bibfield  {author} {\bibinfo {author} {\bibfnamefont {S.}~\bibnamefont
  {Sur}}\ and\ \bibinfo {author} {\bibfnamefont {S.-S.}\ \bibnamefont {Lee}},\
  }\href {\doibase 10.1103/PhysRevB.90.045121} {\bibfield  {journal} {\bibinfo
  {journal} {Phys. Rev. B}\ }\textbf {\bibinfo {volume} {90}},\ \bibinfo
  {pages} {045121} (\bibinfo {year} {2014})}\BibitemShut {NoStop}%
\bibitem [{\citenamefont {Holder}\ and\ \citenamefont
  {Metzner}(2015)}]{PhysRevB.92.041112}%
  \BibitemOpen
  \bibfield  {author} {\bibinfo {author} {\bibfnamefont {T.}~\bibnamefont
  {Holder}}\ and\ \bibinfo {author} {\bibfnamefont {W.}~\bibnamefont
  {Metzner}},\ }\href {\doibase 10.1103/PhysRevB.92.041112} {\bibfield
  {journal} {\bibinfo  {journal} {Phys. Rev. B}\ }\textbf {\bibinfo {volume}
  {92}},\ \bibinfo {pages} {041112} (\bibinfo {year} {2015})}\BibitemShut
  {NoStop}%
\bibitem [{\citenamefont {Oganesyan}\ \emph {et~al.}(2001)\citenamefont
  {Oganesyan}, \citenamefont {Kivelson},\ and\ \citenamefont
  {Fradkin}}]{PhysRevB.64.195109}%
  \BibitemOpen
  \bibfield  {author} {\bibinfo {author} {\bibfnamefont {V.}~\bibnamefont
  {Oganesyan}}, \bibinfo {author} {\bibfnamefont {S.~A.}\ \bibnamefont
  {Kivelson}}, \ and\ \bibinfo {author} {\bibfnamefont {E.}~\bibnamefont
  {Fradkin}},\ }\href {\doibase 10.1103/PhysRevB.64.195109} {\bibfield
  {journal} {\bibinfo  {journal} {Phys. Rev. B}\ }\textbf {\bibinfo {volume}
  {64}},\ \bibinfo {pages} {195109} (\bibinfo {year} {2001})}\BibitemShut
  {NoStop}%
\bibitem [{\citenamefont {Metzner}\ \emph {et~al.}(2003)\citenamefont
  {Metzner}, \citenamefont {Rohe},\ and\ \citenamefont
  {Andergassen}}]{PhysRevLett.91.066402}%
  \BibitemOpen
  \bibfield  {author} {\bibinfo {author} {\bibfnamefont {W.}~\bibnamefont
  {Metzner}}, \bibinfo {author} {\bibfnamefont {D.}~\bibnamefont {Rohe}}, \
  and\ \bibinfo {author} {\bibfnamefont {S.}~\bibnamefont {Andergassen}},\
  }\href {\doibase 10.1103/PhysRevLett.91.066402} {\bibfield  {journal}
  {\bibinfo  {journal} {Phys. Rev. Lett.}\ }\textbf {\bibinfo {volume} {91}},\
  \bibinfo {pages} {066402} (\bibinfo {year} {2003})}\BibitemShut {NoStop}%
\bibitem [{\citenamefont {Lawler}\ and\ \citenamefont
  {Fradkin}(2007)}]{PhysRevB.75.033304}%
  \BibitemOpen
  \bibfield  {author} {\bibinfo {author} {\bibfnamefont {M.~J.}\ \bibnamefont
  {Lawler}}\ and\ \bibinfo {author} {\bibfnamefont {E.}~\bibnamefont
  {Fradkin}},\ }\href {\doibase 10.1103/PhysRevB.75.033304} {\bibfield
  {journal} {\bibinfo  {journal} {Phys. Rev. B}\ }\textbf {\bibinfo {volume}
  {75}},\ \bibinfo {pages} {033304} (\bibinfo {year} {2007})}\BibitemShut
  {NoStop}%
\bibitem [{\citenamefont {Rech}\ \emph {et~al.}(2006)\citenamefont {Rech},
  \citenamefont {P\'epin},\ and\ \citenamefont
  {Chubukov}}]{PhysRevB.74.195126}%
  \BibitemOpen
  \bibfield  {author} {\bibinfo {author} {\bibfnamefont {J.}~\bibnamefont
  {Rech}}, \bibinfo {author} {\bibfnamefont {C.}~\bibnamefont {P\'epin}}, \
  and\ \bibinfo {author} {\bibfnamefont {A.~V.}\ \bibnamefont {Chubukov}},\
  }\href {\doibase 10.1103/PhysRevB.74.195126} {\bibfield  {journal} {\bibinfo
  {journal} {Phys. Rev. B}\ }\textbf {\bibinfo {volume} {74}},\ \bibinfo
  {pages} {195126} (\bibinfo {year} {2006})}\BibitemShut {NoStop}%
\bibitem [{\citenamefont {Maslov}\ and\ \citenamefont
  {Chubukov}(2010)}]{PhysRevB.81.045110}%
  \BibitemOpen
  \bibfield  {author} {\bibinfo {author} {\bibfnamefont {D.~L.}\ \bibnamefont
  {Maslov}}\ and\ \bibinfo {author} {\bibfnamefont {A.~V.}\ \bibnamefont
  {Chubukov}},\ }\href {\doibase 10.1103/PhysRevB.81.045110} {\bibfield
  {journal} {\bibinfo  {journal} {Phys. Rev. B}\ }\textbf {\bibinfo {volume}
  {81}},\ \bibinfo {pages} {045110} (\bibinfo {year} {2010})}\BibitemShut
  {NoStop}%
\bibitem [{\citenamefont {Zacharias}\ \emph {et~al.}(2009)\citenamefont
  {Zacharias}, \citenamefont {W\"olfle},\ and\ \citenamefont
  {Garst}}]{PhysRevB.80.165116}%
  \BibitemOpen
  \bibfield  {author} {\bibinfo {author} {\bibfnamefont {M.}~\bibnamefont
  {Zacharias}}, \bibinfo {author} {\bibfnamefont {P.}~\bibnamefont {W\"olfle}},
  \ and\ \bibinfo {author} {\bibfnamefont {M.}~\bibnamefont {Garst}},\ }\href
  {\doibase 10.1103/PhysRevB.80.165116} {\bibfield  {journal} {\bibinfo
  {journal} {Phys. Rev. B}\ }\textbf {\bibinfo {volume} {80}},\ \bibinfo
  {pages} {165116} (\bibinfo {year} {2009})}\BibitemShut {NoStop}%
\bibitem [{\citenamefont {Lee}(2008)}]{PhysRevB.78.085129}%
  \BibitemOpen
  \bibfield  {author} {\bibinfo {author} {\bibfnamefont {S.-S.}\ \bibnamefont
  {Lee}},\ }\href {\doibase 10.1103/PhysRevB.78.085129} {\bibfield  {journal}
  {\bibinfo  {journal} {Phys. Rev. B}\ }\textbf {\bibinfo {volume} {78}},\
  \bibinfo {pages} {085129} (\bibinfo {year} {2008})}\BibitemShut {NoStop}%
\bibitem [{\citenamefont {{S{\"a}terskog}}\ \emph {et~al.}(2016)\citenamefont
  {{S{\"a}terskog}}, \citenamefont {{Meszena}},\ and\ \citenamefont
  {{Schalm}}}]{2016arXiv161205326S}%
  \BibitemOpen
  \bibfield  {author} {\bibinfo {author} {\bibfnamefont {P.}~\bibnamefont
  {{S{\"a}terskog}}}, \bibinfo {author} {\bibfnamefont {B.}~\bibnamefont
  {{Meszena}}}, \ and\ \bibinfo {author} {\bibfnamefont {K.}~\bibnamefont
  {{Schalm}}},\ }\href@noop {} {\bibfield  {journal} {\bibinfo  {journal}
  {ArXiv e-prints}\ } (\bibinfo {year} {2016})},\ \Eprint
  {http://arxiv.org/abs/1612.05326} {arXiv:1612.05326} \BibitemShut {NoStop}%
\bibitem [{\citenamefont {Fitzpatrick}\ \emph {et~al.}(2014)\citenamefont
  {Fitzpatrick}, \citenamefont {Kachru}, \citenamefont {Kaplan},\ and\
  \citenamefont {Raghu}}]{PhysRevB.89.165114}%
  \BibitemOpen
  \bibfield  {author} {\bibinfo {author} {\bibfnamefont {A.~L.}\ \bibnamefont
  {Fitzpatrick}}, \bibinfo {author} {\bibfnamefont {S.}~\bibnamefont {Kachru}},
  \bibinfo {author} {\bibfnamefont {J.}~\bibnamefont {Kaplan}}, \ and\ \bibinfo
  {author} {\bibfnamefont {S.}~\bibnamefont {Raghu}},\ }\href {\doibase
  10.1103/PhysRevB.89.165114} {\bibfield  {journal} {\bibinfo  {journal} {Phys.
  Rev. B}\ }\textbf {\bibinfo {volume} {89}},\ \bibinfo {pages} {165114}
  (\bibinfo {year} {2014})}\BibitemShut {NoStop}%
\bibitem [{\citenamefont {Mandal}\ and\ \citenamefont
  {Lee}(2015)}]{PhysRevB.92.035141}%
  \BibitemOpen
  \bibfield  {author} {\bibinfo {author} {\bibfnamefont {I.}~\bibnamefont
  {Mandal}}\ and\ \bibinfo {author} {\bibfnamefont {S.-S.}\ \bibnamefont
  {Lee}},\ }\href {\doibase 10.1103/PhysRevB.92.035141} {\bibfield  {journal}
  {\bibinfo  {journal} {Phys. Rev. B}\ }\textbf {\bibinfo {volume} {92}},\
  \bibinfo {pages} {035141} (\bibinfo {year} {2015})}\BibitemShut {NoStop}%
\bibitem [{\citenamefont {Chakravarty}\ \emph {et~al.}(1995)\citenamefont
  {Chakravarty}, \citenamefont {Norton},\ and\ \citenamefont
  {Sylju\aa{}sen}}]{PhysRevLett.74.1423}%
  \BibitemOpen
  \bibfield  {author} {\bibinfo {author} {\bibfnamefont {S.}~\bibnamefont
  {Chakravarty}}, \bibinfo {author} {\bibfnamefont {R.~E.}\ \bibnamefont
  {Norton}}, \ and\ \bibinfo {author} {\bibfnamefont {O.~F.}\ \bibnamefont
  {Sylju\aa{}sen}},\ }\href {\doibase 10.1103/PhysRevLett.74.1423} {\bibfield
  {journal} {\bibinfo  {journal} {Phys. Rev. Lett.}\ }\textbf {\bibinfo
  {volume} {74}},\ \bibinfo {pages} {1423} (\bibinfo {year}
  {1995})}\BibitemShut {NoStop}%
\bibitem [{\citenamefont {Fitzpatrick}\ \emph {et~al.}(2013)\citenamefont
  {Fitzpatrick}, \citenamefont {Kachru}, \citenamefont {Kaplan},\ and\
  \citenamefont {Raghu}}]{PhysRevB.88.125116}%
  \BibitemOpen
  \bibfield  {author} {\bibinfo {author} {\bibfnamefont {A.~L.}\ \bibnamefont
  {Fitzpatrick}}, \bibinfo {author} {\bibfnamefont {S.}~\bibnamefont {Kachru}},
  \bibinfo {author} {\bibfnamefont {J.}~\bibnamefont {Kaplan}}, \ and\ \bibinfo
  {author} {\bibfnamefont {S.}~\bibnamefont {Raghu}},\ }\href {\doibase
  10.1103/PhysRevB.88.125116} {\bibfield  {journal} {\bibinfo  {journal} {Phys.
  Rev. B}\ }\textbf {\bibinfo {volume} {88}},\ \bibinfo {pages} {125116}
  (\bibinfo {year} {2013})}\BibitemShut {NoStop}%
\bibitem [{\citenamefont {Senthil}\ and\ \citenamefont
  {Shankar}(2009)}]{PhysRevLett.102.046406}%
  \BibitemOpen
  \bibfield  {author} {\bibinfo {author} {\bibfnamefont {T.}~\bibnamefont
  {Senthil}}\ and\ \bibinfo {author} {\bibfnamefont {R.}~\bibnamefont
  {Shankar}},\ }\href {\doibase 10.1103/PhysRevLett.102.046406} {\bibfield
  {journal} {\bibinfo  {journal} {Phys. Rev. Lett.}\ }\textbf {\bibinfo
  {volume} {102}},\ \bibinfo {pages} {046406} (\bibinfo {year}
  {2009})}\BibitemShut {NoStop}%
\bibitem [{\citenamefont {Sur}\ and\ \citenamefont
  {Lee}(2015)}]{PhysRevB.91.125136}%
  \BibitemOpen
  \bibfield  {author} {\bibinfo {author} {\bibfnamefont {S.}~\bibnamefont
  {Sur}}\ and\ \bibinfo {author} {\bibfnamefont {S.-S.}\ \bibnamefont {Lee}},\
  }\href {\doibase 10.1103/PhysRevB.91.125136} {\bibfield  {journal} {\bibinfo
  {journal} {Phys. Rev. B}\ }\textbf {\bibinfo {volume} {91}},\ \bibinfo
  {pages} {125136} (\bibinfo {year} {2015})}\BibitemShut {NoStop}%
\bibitem [{\citenamefont {Eberlein}\ \emph {et~al.}(2017)\citenamefont
  {Eberlein}, \citenamefont {Patel},\ and\ \citenamefont
  {Sachdev}}]{PhysRevB.95.075127}%
  \BibitemOpen
  \bibfield  {author} {\bibinfo {author} {\bibfnamefont {A.}~\bibnamefont
  {Eberlein}}, \bibinfo {author} {\bibfnamefont {A.~A.}\ \bibnamefont {Patel}},
  \ and\ \bibinfo {author} {\bibfnamefont {S.}~\bibnamefont {Sachdev}},\ }\href
  {\doibase 10.1103/PhysRevB.95.075127} {\bibfield  {journal} {\bibinfo
  {journal} {Phys. Rev. B}\ }\textbf {\bibinfo {volume} {95}},\ \bibinfo
  {pages} {075127} (\bibinfo {year} {2017})}\BibitemShut {NoStop}%
\bibitem [{\citenamefont {Son}(1999)}]{PhysRevD.59.094019}%
  \BibitemOpen
  \bibfield  {author} {\bibinfo {author} {\bibfnamefont {D.~T.}\ \bibnamefont
  {Son}},\ }\href {\doibase 10.1103/PhysRevD.59.094019} {\bibfield  {journal}
  {\bibinfo  {journal} {Phys. Rev. D}\ }\textbf {\bibinfo {volume} {59}},\
  \bibinfo {pages} {094019} (\bibinfo {year} {1999})}\BibitemShut {NoStop}%
\bibitem [{\citenamefont {Metlitski}\ \emph {et~al.}(2015)\citenamefont
  {Metlitski}, \citenamefont {Mross}, \citenamefont {Sachdev},\ and\
  \citenamefont {Senthil}}]{PhysRevB.91.115111}%
  \BibitemOpen
  \bibfield  {author} {\bibinfo {author} {\bibfnamefont {M.~A.}\ \bibnamefont
  {Metlitski}}, \bibinfo {author} {\bibfnamefont {D.~F.}\ \bibnamefont
  {Mross}}, \bibinfo {author} {\bibfnamefont {S.}~\bibnamefont {Sachdev}}, \
  and\ \bibinfo {author} {\bibfnamefont {T.}~\bibnamefont {Senthil}},\ }\href
  {\doibase 10.1103/PhysRevB.91.115111} {\bibfield  {journal} {\bibinfo
  {journal} {Phys. Rev. B}\ }\textbf {\bibinfo {volume} {91}},\ \bibinfo
  {pages} {115111} (\bibinfo {year} {2015})}\BibitemShut {NoStop}%
\bibitem [{\citenamefont {Raghu}\ \emph {et~al.}(2015)\citenamefont {Raghu},
  \citenamefont {Torroba},\ and\ \citenamefont {Wang}}]{PhysRevB.92.205104}%
  \BibitemOpen
  \bibfield  {author} {\bibinfo {author} {\bibfnamefont {S.}~\bibnamefont
  {Raghu}}, \bibinfo {author} {\bibfnamefont {G.}~\bibnamefont {Torroba}}, \
  and\ \bibinfo {author} {\bibfnamefont {H.}~\bibnamefont {Wang}},\ }\href
  {\doibase 10.1103/PhysRevB.92.205104} {\bibfield  {journal} {\bibinfo
  {journal} {Phys. Rev. B}\ }\textbf {\bibinfo {volume} {92}},\ \bibinfo
  {pages} {205104} (\bibinfo {year} {2015})}\BibitemShut {NoStop}%
\bibitem [{\citenamefont {Lederer}\ \emph {et~al.}(2015)\citenamefont
  {Lederer}, \citenamefont {Schattner}, \citenamefont {Berg},\ and\
  \citenamefont {Kivelson}}]{PhysRevLett.114.097001}%
  \BibitemOpen
  \bibfield  {author} {\bibinfo {author} {\bibfnamefont {S.}~\bibnamefont
  {Lederer}}, \bibinfo {author} {\bibfnamefont {Y.}~\bibnamefont {Schattner}},
  \bibinfo {author} {\bibfnamefont {E.}~\bibnamefont {Berg}}, \ and\ \bibinfo
  {author} {\bibfnamefont {S.~A.}\ \bibnamefont {Kivelson}},\ }\href {\doibase
  10.1103/PhysRevLett.114.097001} {\bibfield  {journal} {\bibinfo  {journal}
  {Phys. Rev. Lett.}\ }\textbf {\bibinfo {volume} {114}},\ \bibinfo {pages}
  {097001} (\bibinfo {year} {2015})}\BibitemShut {NoStop}%
\bibitem [{\citenamefont {Mandal}(2016)}]{PhysRevB.94.115138}%
  \BibitemOpen
  \bibfield  {author} {\bibinfo {author} {\bibfnamefont {I.}~\bibnamefont
  {Mandal}},\ }\href {\doibase 10.1103/PhysRevB.94.115138} {\bibfield
  {journal} {\bibinfo  {journal} {Phys. Rev. B}\ }\textbf {\bibinfo {volume}
  {94}},\ \bibinfo {pages} {115138} (\bibinfo {year} {2016})}\BibitemShut
  {NoStop}%
\bibitem [{\citenamefont {{Lederer}}\ \emph {et~al.}(2016)\citenamefont
  {{Lederer}}, \citenamefont {{Schattner}}, \citenamefont {{Berg}},\ and\
  \citenamefont {{Kivelson}}}]{2016arXiv161201542L}%
  \BibitemOpen
  \bibfield  {author} {\bibinfo {author} {\bibfnamefont {S.}~\bibnamefont
  {{Lederer}}}, \bibinfo {author} {\bibfnamefont {Y.}~\bibnamefont
  {{Schattner}}}, \bibinfo {author} {\bibfnamefont {E.}~\bibnamefont {{Berg}}},
  \ and\ \bibinfo {author} {\bibfnamefont {S.~A.}\ \bibnamefont {{Kivelson}}},\
  }\href@noop {} {\bibfield  {journal} {\bibinfo  {journal} {ArXiv e-prints}\ }
  (\bibinfo {year} {2016})},\ \Eprint {http://arxiv.org/abs/1612.01542}
  {arXiv:1612.01542 [cond-mat.str-el]} \BibitemShut {NoStop}%
\bibitem [{\citenamefont {Balents}\ and\ \citenamefont
  {Fisher}(1996)}]{PhysRevLett.76.2782}%
  \BibitemOpen
  \bibfield  {author} {\bibinfo {author} {\bibfnamefont {L.}~\bibnamefont
  {Balents}}\ and\ \bibinfo {author} {\bibfnamefont {M.~P.~A.}\ \bibnamefont
  {Fisher}},\ }\href {\doibase 10.1103/PhysRevLett.76.2782} {\bibfield
  {journal} {\bibinfo  {journal} {Phys. Rev. Lett.}\ }\textbf {\bibinfo
  {volume} {76}},\ \bibinfo {pages} {2782} (\bibinfo {year}
  {1996})}\BibitemShut {NoStop}%
\bibitem [{\citenamefont {Wen}(1990)}]{PhysRevB.41.12838}%
  \BibitemOpen
  \bibfield  {author} {\bibinfo {author} {\bibfnamefont {X.~G.}\ \bibnamefont
  {Wen}},\ }\href {\doibase 10.1103/PhysRevB.41.12838} {\bibfield  {journal}
  {\bibinfo  {journal} {Phys. Rev. B}\ }\textbf {\bibinfo {volume} {41}},\
  \bibinfo {pages} {12838} (\bibinfo {year} {1990})}\BibitemShut {NoStop}%
\bibitem [{\citenamefont {Patel}\ and\ \citenamefont
  {Sachdev}(2017)}]{Patel21022017}%
  \BibitemOpen
  \bibfield  {author} {\bibinfo {author} {\bibfnamefont {A.~A.}\ \bibnamefont
  {Patel}}\ and\ \bibinfo {author} {\bibfnamefont {S.}~\bibnamefont
  {Sachdev}},\ }\href {\doibase 10.1073/pnas.1618185114} {\bibfield  {journal}
  {\bibinfo  {journal} {Proceedings of the National Academy of Sciences}\
  }\textbf {\bibinfo {volume} {114}},\ \bibinfo {pages} {1844} (\bibinfo {year}
  {2017})},\ \Eprint
  {http://arxiv.org/abs/http://www.pnas.org/content/114/8/1844.full.pdf}
  {http://www.pnas.org/content/114/8/1844.full.pdf} \BibitemShut {NoStop}%
\bibitem [{\citenamefont {Abanov}\ and\ \citenamefont
  {Chubukov}(2000)}]{PhysRevLett.84.5608}%
  \BibitemOpen
  \bibfield  {author} {\bibinfo {author} {\bibfnamefont {A.}~\bibnamefont
  {Abanov}}\ and\ \bibinfo {author} {\bibfnamefont {A.~V.}\ \bibnamefont
  {Chubukov}},\ }\href {\doibase 10.1103/PhysRevLett.84.5608} {\bibfield
  {journal} {\bibinfo  {journal} {Phys. Rev. Lett.}\ }\textbf {\bibinfo
  {volume} {84}},\ \bibinfo {pages} {5608} (\bibinfo {year}
  {2000})}\BibitemShut {NoStop}%
\bibitem [{\citenamefont {Abanov}\ and\ \citenamefont
  {Chubukov}(2004)}]{PhysRevLett.93.255702}%
  \BibitemOpen
  \bibfield  {author} {\bibinfo {author} {\bibfnamefont {A.}~\bibnamefont
  {Abanov}}\ and\ \bibinfo {author} {\bibfnamefont {A.}~\bibnamefont
  {Chubukov}},\ }\href {\doibase 10.1103/PhysRevLett.93.255702} {\bibfield
  {journal} {\bibinfo  {journal} {Phys. Rev. Lett.}\ }\textbf {\bibinfo
  {volume} {93}},\ \bibinfo {pages} {255702} (\bibinfo {year}
  {2004})}\BibitemShut {NoStop}%
\bibitem [{\citenamefont {Metlitski}\ and\ \citenamefont
  {Sachdev}(2010{\natexlab{b}})}]{PhysRevB.82.075128}%
  \BibitemOpen
  \bibfield  {author} {\bibinfo {author} {\bibfnamefont {M.~A.}\ \bibnamefont
  {Metlitski}}\ and\ \bibinfo {author} {\bibfnamefont {S.}~\bibnamefont
  {Sachdev}},\ }\href {\doibase 10.1103/PhysRevB.82.075128} {\bibfield
  {journal} {\bibinfo  {journal} {Phys. Rev. B}\ }\textbf {\bibinfo {volume}
  {82}},\ \bibinfo {pages} {075128} (\bibinfo {year}
  {2010}{\natexlab{b}})}\BibitemShut {NoStop}%
\bibitem [{\citenamefont {Schlief}\ \emph {et~al.}()\citenamefont {Schlief},
  \citenamefont {Lunts},\ and\ \citenamefont {Lee}}]{SDWgenD}%
  \BibitemOpen
  \bibfield  {author} {\bibinfo {author} {\bibfnamefont {A.}~\bibnamefont
  {Schlief}}, \bibinfo {author} {\bibfnamefont {P.}~\bibnamefont {Lunts}}, \
  and\ \bibinfo {author} {\bibfnamefont {S.-S.}\ \bibnamefont {Lee}},\
  }\href@noop {} {\bibinfo  {journal} {in preparation}\ }\BibitemShut {NoStop}%
\bibitem [{\citenamefont {{Schlief}}\ \emph {et~al.}(2016)\citenamefont
  {{Schlief}}, \citenamefont {{Lunts}},\ and\ \citenamefont
  {{Lee}}}]{2016arXiv160806927S}%
  \BibitemOpen
\bibfield  {journal} {  }\bibfield  {author} {\bibinfo {author} {\bibfnamefont
  {A.}~\bibnamefont {{Schlief}}}, \bibinfo {author} {\bibfnamefont
  {P.}~\bibnamefont {{Lunts}}}, \ and\ \bibinfo {author} {\bibfnamefont
  {S.-S.}\ \bibnamefont {{Lee}}},\ }\href@noop {} {\bibfield  {journal}
  {\bibinfo  {journal} {ArXiv e-prints}\ } (\bibinfo {year} {2016})},\ \Eprint
  {http://arxiv.org/abs/1608.06927} {arXiv:1608.06927 [cond-mat.str-el]}
  \BibitemShut {NoStop}%
\bibitem [{\citenamefont {{Lunts}}\ \emph {et~al.}(2017)\citenamefont
  {{Lunts}}, \citenamefont {{Schlief}},\ and\ \citenamefont
  {{Lee}}}]{2017arXiv170108218L}%
  \BibitemOpen
  \bibfield  {author} {\bibinfo {author} {\bibfnamefont {P.}~\bibnamefont
  {{Lunts}}}, \bibinfo {author} {\bibfnamefont {A.}~\bibnamefont {{Schlief}}},
  \ and\ \bibinfo {author} {\bibfnamefont {S.-S.}\ \bibnamefont {{Lee}}},\
  }\href@noop {} {\bibfield  {journal} {\bibinfo  {journal} {ArXiv e-prints}\ }
  (\bibinfo {year} {2017})},\ \Eprint {http://arxiv.org/abs/1701.08218}
  {arXiv:1701.08218 [cond-mat.str-el]} \BibitemShut {NoStop}%
\bibitem [{\citenamefont {Sur}\ and\ \citenamefont
  {Lee}(2016)}]{PhysRevB.94.195135}%
  \BibitemOpen
  \bibfield  {author} {\bibinfo {author} {\bibfnamefont {S.}~\bibnamefont
  {Sur}}\ and\ \bibinfo {author} {\bibfnamefont {S.-S.}\ \bibnamefont {Lee}},\
  }\href {\doibase 10.1103/PhysRevB.94.195135} {\bibfield  {journal} {\bibinfo
  {journal} {Phys. Rev. B}\ }\textbf {\bibinfo {volume} {94}},\ \bibinfo
  {pages} {195135} (\bibinfo {year} {2016})}\BibitemShut {NoStop}%
\bibitem [{\citenamefont {Huh}\ and\ \citenamefont
  {Sachdev}(2008)}]{PhysRevB.78.064512}%
  \BibitemOpen
  \bibfield  {author} {\bibinfo {author} {\bibfnamefont {Y.}~\bibnamefont
  {Huh}}\ and\ \bibinfo {author} {\bibfnamefont {S.}~\bibnamefont {Sachdev}},\
  }\href {\doibase 10.1103/PhysRevB.78.064512} {\bibfield  {journal} {\bibinfo
  {journal} {Phys. Rev. B}\ }\textbf {\bibinfo {volume} {78}},\ \bibinfo
  {pages} {064512} (\bibinfo {year} {2008})}\BibitemShut {NoStop}%
\bibitem [{\citenamefont {Abanov}\ \emph {et~al.}(2003)\citenamefont {Abanov},
  \citenamefont {Chubukov},\ and\ \citenamefont
  {Schmalian}}]{doi:10.1080/0001873021000057123}%
  \BibitemOpen
  \bibfield  {author} {\bibinfo {author} {\bibfnamefont {A.}~\bibnamefont
  {Abanov}}, \bibinfo {author} {\bibfnamefont {A.~V.}\ \bibnamefont
  {Chubukov}}, \ and\ \bibinfo {author} {\bibfnamefont {J.}~\bibnamefont
  {Schmalian}},\ }\href {\doibase 10.1080/0001873021000057123} {\bibfield
  {journal} {\bibinfo  {journal} {Advances in Physics}\ }\textbf {\bibinfo
  {volume} {52}},\ \bibinfo {pages} {119} (\bibinfo {year} {2003})},\ \Eprint
  {http://arxiv.org/abs/http://www.tandfonline.com/doi/pdf/10.1080/0001873021000057123}
  {http://www.tandfonline.com/doi/pdf/10.1080/0001873021000057123} \BibitemShut
  {NoStop}%
\bibitem [{\citenamefont {Maier}\ and\ \citenamefont
  {Strack}(2016)}]{PhysRevB.93.165114}%
  \BibitemOpen
  \bibfield  {author} {\bibinfo {author} {\bibfnamefont {S.~A.}\ \bibnamefont
  {Maier}}\ and\ \bibinfo {author} {\bibfnamefont {P.}~\bibnamefont {Strack}},\
  }\href {\doibase 10.1103/PhysRevB.93.165114} {\bibfield  {journal} {\bibinfo
  {journal} {Phys. Rev. B}\ }\textbf {\bibinfo {volume} {93}},\ \bibinfo
  {pages} {165114} (\bibinfo {year} {2016})}\BibitemShut {NoStop}%
\bibitem [{\citenamefont {Scalapino}\ \emph {et~al.}(1986)\citenamefont
  {Scalapino}, \citenamefont {Loh~Jr},\ and\ \citenamefont
  {Hirsch}}]{scalapino1986d}%
  \BibitemOpen
  \bibfield  {author} {\bibinfo {author} {\bibfnamefont {D.}~\bibnamefont
  {Scalapino}}, \bibinfo {author} {\bibfnamefont {E.}~\bibnamefont {Loh~Jr}}, \
  and\ \bibinfo {author} {\bibfnamefont {J.}~\bibnamefont {Hirsch}},\
  }\href@noop {} {\bibfield  {journal} {\bibinfo  {journal} {Physical Review
  B}\ }\textbf {\bibinfo {volume} {34}},\ \bibinfo {pages} {8190} (\bibinfo
  {year} {1986})}\BibitemShut {NoStop}%
\bibitem [{\citenamefont {Miyake}\ \emph {et~al.}(1986)\citenamefont {Miyake},
  \citenamefont {Schmitt-Rink},\ and\ \citenamefont
  {Varma}}]{PhysRevB.34.6554}%
  \BibitemOpen
  \bibfield  {author} {\bibinfo {author} {\bibfnamefont {K.}~\bibnamefont
  {Miyake}}, \bibinfo {author} {\bibfnamefont {S.}~\bibnamefont
  {Schmitt-Rink}}, \ and\ \bibinfo {author} {\bibfnamefont {C.~M.}\
  \bibnamefont {Varma}},\ }\href {\doibase 10.1103/PhysRevB.34.6554} {\bibfield
   {journal} {\bibinfo  {journal} {Phys. Rev. B}\ }\textbf {\bibinfo {volume}
  {34}},\ \bibinfo {pages} {6554} (\bibinfo {year} {1986})}\BibitemShut
  {NoStop}%
\bibitem [{\citenamefont {Moriya}\ \emph {et~al.}(1990)\citenamefont {Moriya},
  \citenamefont {Takahashi},\ and\ \citenamefont
  {Ueda}}]{doi:10.1143/JPSJ.59.2905}%
  \BibitemOpen
  \bibfield  {author} {\bibinfo {author} {\bibfnamefont {T.}~\bibnamefont
  {Moriya}}, \bibinfo {author} {\bibfnamefont {Y.}~\bibnamefont {Takahashi}}, \
  and\ \bibinfo {author} {\bibfnamefont {K.}~\bibnamefont {Ueda}},\ }\href
  {\doibase 10.1143/JPSJ.59.2905} {\bibfield  {journal} {\bibinfo  {journal}
  {Journal of the Physical Society of Japan}\ }\textbf {\bibinfo {volume}
  {59}},\ \bibinfo {pages} {2905} (\bibinfo {year} {1990})},\ \Eprint
  {http://arxiv.org/abs/http://dx.doi.org/10.1143/JPSJ.59.2905}
  {http://dx.doi.org/10.1143/JPSJ.59.2905} \BibitemShut {NoStop}%
\bibitem [{\citenamefont {Berg}\ \emph {et~al.}(2012)\citenamefont {Berg},
  \citenamefont {Metlitski},\ and\ \citenamefont {Sachdev}}]{Berg21122012}%
  \BibitemOpen
  \bibfield  {author} {\bibinfo {author} {\bibfnamefont {E.}~\bibnamefont
  {Berg}}, \bibinfo {author} {\bibfnamefont {M.~A.}\ \bibnamefont {Metlitski}},
  \ and\ \bibinfo {author} {\bibfnamefont {S.}~\bibnamefont {Sachdev}},\ }\href
  {\doibase 10.1126/science.1227769} {\bibfield  {journal} {\bibinfo  {journal}
  {Science}\ }\textbf {\bibinfo {volume} {338}},\ \bibinfo {pages} {1606}
  (\bibinfo {year} {2012})},\ \Eprint
  {http://arxiv.org/abs/http://www.sciencemag.org/content/338/6114/1606.full.pdf}
  {http://www.sciencemag.org/content/338/6114/1606.full.pdf} \BibitemShut
  {NoStop}%
\bibitem [{\citenamefont {{Li}}\ \emph {et~al.}(2015)\citenamefont {{Li}},
  \citenamefont {{Wang}}, \citenamefont {{Yao}},\ and\ \citenamefont
  {{Lee}}}]{2015arXiv151204541L}%
  \BibitemOpen
  \bibfield  {author} {\bibinfo {author} {\bibfnamefont {Z.-X.}\ \bibnamefont
  {{Li}}}, \bibinfo {author} {\bibfnamefont {F.}~\bibnamefont {{Wang}}},
  \bibinfo {author} {\bibfnamefont {H.}~\bibnamefont {{Yao}}}, \ and\ \bibinfo
  {author} {\bibfnamefont {D.-H.}\ \bibnamefont {{Lee}}},\ }\href@noop {}
  {\bibfield  {journal} {\bibinfo  {journal} {ArXiv e-prints}\ } (\bibinfo
  {year} {2015})},\ \Eprint {http://arxiv.org/abs/1512.04541} {arXiv:1512.04541
  [cond-mat.supr-con]} \BibitemShut {NoStop}%
\bibitem [{\citenamefont {Schattner}\ \emph {et~al.}(2016)\citenamefont
  {Schattner}, \citenamefont {Gerlach}, \citenamefont {Trebst},\ and\
  \citenamefont {Berg}}]{PhysRevLett.117.097002}%
  \BibitemOpen
  \bibfield  {author} {\bibinfo {author} {\bibfnamefont {Y.}~\bibnamefont
  {Schattner}}, \bibinfo {author} {\bibfnamefont {M.~H.}\ \bibnamefont
  {Gerlach}}, \bibinfo {author} {\bibfnamefont {S.}~\bibnamefont {Trebst}}, \
  and\ \bibinfo {author} {\bibfnamefont {E.}~\bibnamefont {Berg}},\ }\href
  {\doibase 10.1103/PhysRevLett.117.097002} {\bibfield  {journal} {\bibinfo
  {journal} {Phys. Rev. Lett.}\ }\textbf {\bibinfo {volume} {117}},\ \bibinfo
  {pages} {097002} (\bibinfo {year} {2016})}\BibitemShut {NoStop}%
\bibitem [{\citenamefont {Monthoux}\ \emph {et~al.}(1992)\citenamefont
  {Monthoux}, \citenamefont {Balatsky},\ and\ \citenamefont
  {Pines}}]{PhysRevB.46.14803}%
  \BibitemOpen
  \bibfield  {author} {\bibinfo {author} {\bibfnamefont {P.}~\bibnamefont
  {Monthoux}}, \bibinfo {author} {\bibfnamefont {A.~V.}\ \bibnamefont
  {Balatsky}}, \ and\ \bibinfo {author} {\bibfnamefont {D.}~\bibnamefont
  {Pines}},\ }\href {\doibase 10.1103/PhysRevB.46.14803} {\bibfield  {journal}
  {\bibinfo  {journal} {Phys. Rev. B}\ }\textbf {\bibinfo {volume} {46}},\
  \bibinfo {pages} {14803} (\bibinfo {year} {1992})}\BibitemShut {NoStop}%
\bibitem [{\citenamefont {Chubukov}\ and\ \citenamefont
  {Schmalian}(2005)}]{PhysRevB.72.174520}%
  \BibitemOpen
  \bibfield  {author} {\bibinfo {author} {\bibfnamefont {A.~V.}\ \bibnamefont
  {Chubukov}}\ and\ \bibinfo {author} {\bibfnamefont {J.}~\bibnamefont
  {Schmalian}},\ }\href {\doibase 10.1103/PhysRevB.72.174520} {\bibfield
  {journal} {\bibinfo  {journal} {Phys. Rev. B}\ }\textbf {\bibinfo {volume}
  {72}},\ \bibinfo {pages} {174520} (\bibinfo {year} {2005})}\BibitemShut
  {NoStop}%
\end{thebibliography}%

\bibliographystyle{apsrev4-1}

\end{document}